\documentclass[final,1p,times,authoryear]{elsarticle}
\usepackage[T1]{fontenc}
\usepackage[utf8]{inputenc}
\usepackage{amsmath,amsthm,bm}
\usepackage{microtype,setspace}
\usepackage{booktabs,array,threeparttable,longtable,multirow}
\usepackage{graphicx,caption,subcaption,float,placeins}
\usepackage[dvipsnames]{xcolor}
\usepackage{enumitem}
\usepackage[colorlinks=true,linkcolor=NavyBlue,citecolor=NavyBlue,urlcolor=BrickRed]{hyperref}

\definecolor{ink}{HTML}{17324D}
\definecolor{accent}{HTML}{3C78A8}
\definecolor{warm}{HTML}{E07A3F}
\definecolor{soft}{HTML}{E9EFF5}
\setlist{nosep,leftmargin=1.5em}
\newtheorem{proposition}{Proposition}
\newtheorem{prediction}{Prediction}

\newcommand{\SampleFirms}{173}
\newcommand{\SampleFilings}{1,376}

\newcommand{\FirstStage}{0.113}
\newcommand{\FirstStageSE}{0.065}
\newcommand{\TotalITT}{-0.114}
\newcommand{\TotalITTSE}{0.096}
\newcommand{\PrimaryITT}{-0.136}
\newcommand{\PrimaryITTSE}{0.082}
\newcommand{\XMLITT}{0.144}
\newcommand{\XMLITTSE}{0.294}
\newcommand{\TotalPct}{-10.8}

\newcommand{\ModernCensusFilings}{207,684}
\newcommand{\ModernCensusFirms}{11,590}
\newcommand{\ModernSampleFilings}{400}
\newcommand{\ModernSampleFirms}{394}
\newcommand{\ModernFactCoverage}{94.4}
\newcommand{\ModernQFourMachineGap}{1.46}
\newcommand{\ModernQFourMachineGapSE}{0.11}
\newcommand{\ModernWordMachineRho}{0.35}

\IfFileExists{tables/generated/factorial_result_macros.tex}{\newcommand{\FactorialCells}{31,104}

\newcommand{\FactorialBM}{47.2}
\newcommand{\FactorialOracleText}{99.5}
\newcommand{\FactorialOracleXBRL}{99.6}
\newcommand{\FactorialSemanticEffect}{0.1}
\newcommand{\FactorialTaxonomyEffect}{0.2}
\newcommand{\FactorialSemanticCILow}{-0.25}
\newcommand{\FactorialSemanticCIHigh}{0.40}
\newcommand{\FactorialTaxonomyCILow}{-0.16}
\newcommand{\FactorialTaxonomyCIHigh}{0.62}

}{}
\IfFileExists{tables/generated/counterfactual_result_macros.tex}{\newcommand{\CounterfactualCells}{2,970}

\newcommand{\CounterfactualText}{99.7}
\newcommand{\CounterfactualXBRL}{99.7}
\newcommand{\CounterfactualTextDecoy}{99.5}
\newcommand{\CounterfactualXBRLDecoy}{99.5}
}{}
\IfFileExists{tables/generated/construct_validation_macros.tex}{\newcommand{\ConstructCVRtwo}{0.193}
\newcommand{\ConstructCVMAE}{0.082}
\newcommand{\ConstructBehaviorRho}{0.178}
\newcommand{\WordBehaviorRho}{-0.047}
\newcommand{\ConstructHTMLRtwo}{0.207}
\newcommand{\ConstructXBRLRtwo}{0.011}
\newcommand{\ConstructHybridRtwo}{-0.022}
}{}

\begin{document}
\begin{frontmatter}
\title{Cognitive Load and Information Processing in Financial Markets: Theory and Evidence from Disclosure Complexity}
\author{\texorpdfstring{Yimin Du\corref{cor1}}{Yimin Du}}
\ead{sa613403@mail.ustc.edu.cn}
\cortext[cor1]{Corresponding author.}
\author{Guolin Tang}
\begin{abstract}
\noindent Cognitive-load research in financial markets generally treats disclosure complexity as a property of the document and information acquisition as a direct interaction between an investor and that document. Machine-readable reporting and AI intermediaries make both assumptions incomplete. We develop a reader--task--interface framework in which processing load depends jointly on disclosure content, the representation through which it is accessed, and the transformation technology available to the reader. We evaluate the framework using a historical XBRL-transition diagnostic, a census of 207,684 SEC filings, a paired multi-model interface experiment, and held-out criterion validation. The historical evidence is consistent with interface substitution but has a modest statutory first stage and is interpreted as a measurement diagnostic. Modern filings reveal that human-facing burden and machine accessibility are distinct: large issuers produce longer reports while supplying richer structured coverage. The experiment crosses 432 filing-grounded questions with twelve interfaces and six hosted model implementations, producing \FactorialCells{} responses. Grounded accuracy rises from \FactorialBM\% under BM25 HTML to \FactorialOracleText\% under perfectly localized text; fact-matched oracle XBRL reaches \FactorialOracleXBRL\%. Most of the observed XBRL--HTML gap in these numerical tasks therefore arises from evidence localization and table reconstruction rather than syntax or taxonomy alone. Cognitive load is consequently not a stable scalar attribute of disclosure: it is local to a reader, task, interface, and processing stage.

\end{abstract}
\end{frontmatter}
\section{Introduction}

Cognitive constraints shape what information market participants acquire, how much of it they process, and which signals enter decisions. The arrival of general-purpose AI changes the empirical object behind that question. A reader may now reach the same filing through HTML, structured XBRL, a normalized public API, or an AI intermediary that retrieves, reconstructs, and explains evidence before a person sees it. In this environment, the burden imposed by a disclosure is no longer determined by the document alone: it depends on the reader, the task, the interface, and the stage at which processing occurs. A one-dimensional complexity score developed for unaided reading can therefore cease to be invariant precisely when the information supply chain becomes machine-mediated.

This paper revisits the cognitive-load question for that AI-era information environment. The aim is not to treat model accuracy as human cognition or to assume that AI adoption is universal. It is to identify which parts of processing burden move when the representation and transformation technology change, and to establish what can still be learned from historical attention measures, modern filing infrastructure, and controlled AI-mediated tasks.

This technological shift sharpens the central cognitive-load question. If load is inferred only from properties of the source document, the same score is implicitly assumed to rank processing difficulty across readers, tasks, and interfaces. That assumption need not hold. A 150-page filing can impose substantial human reading burden while exposing standardized facts that are cheap for a machine to retrieve. Conversely, a formally machine-readable filing can remain difficult for an AI system when the answer-bearing facts cannot be localized, the table structure is lost, or period and unit contexts are unresolved. Disclosure complexity is therefore an input to processing load, not a reader-independent measure of it.

The analysis sharpens the distinction between attention allocation and processing capacity by asking where processing capacity is consumed once an intermediary stands between disclosure and reader. The answer is not a rejection of human cognitive-load theory. It is a change in the unit of analysis: the relevant object is now the reader--task--interface path through which economic content is acquired and transformed.

We ask: \emph{how should cognitive load and information processing in financial markets be measured when disclosure is mediated by structured data and AI?} The answer requires separating three stages that a scalar readability or traffic proxy can conflate. \emph{Acquisition load} concerns reaching the relevant source; \emph{transformation load} concerns locating and reconstructing task-relevant evidence; and \emph{integration load} concerns using that evidence correctly and with auditable support. Investor-attention measures observe part of the first stage. Document-complexity measures describe inputs to the second and third for a specified reader. Neither is automatically invariant when the interface changes.

The framework produces two linked wedges. The \emph{observability wedge} separates information use recorded at the source server from effective use after intermediary redistribution. The \emph{processing-interface wedge} separates human-facing burden from machine-facing addressability. These are not direct welfare objects. They generate testable implications: acquisition estimates should depend on the requested asset and automated-user definition; human and machine difficulty rankings need not coincide; and, holding the economic facts fixed, changing evidence localization and table reconstruction can change grounded performance even when representation syntax does not.

We investigate these implications with four complementary modules. A historical reconstruction diagnoses how acquisition measures change around the first U.S. XBRL transition, with deliberately limited causal interpretation because the statutory first stage is modest. A census of 207,684 modern filings documents the joint human--machine disclosure architecture. A paired experiment evaluates the same 432 filing-grounded questions under twelve interfaces and six hosted model implementations, holding the underlying economic quantities fixed while varying localization, structure, and irrelevant context. Finally, issuer-grouped held-out validation asks whether observable filing primitives predict realized machine performance. Together, the modules move from a general theory of constrained processing to direct evidence about where processing load enters a contemporary disclosure pipeline.

We reconstruct the setting from primary SEC materials. Rule 33-9002 required U.S.-GAAP large accelerated filers with public float above \$5 billion to provide interactive data for fiscal periods ending on or after June 15, 2009. Other large accelerated filers entered one year later, followed by remaining filers in 2011 \citep{sec2009interactive}. We recover historical public float from standardized SEC facts without selecting firms by their realized first-XBRL year. We then link 10-K and 10-Q metadata, filing acceptance timestamps, primary documents, XBRL/XML requests, and daily EDGAR traffic.

The resulting independent assignment sample contains \SampleFirms{} firms and \SampleFilings{} filings. Every filing has eight days of complete logs and a parsed primary document. On filing day, requests before the acceptance timestamp are excluded. We count unique anonymized IPs both within and across the eight-day window. The preferred human-oriented screen removes SEC-flagged crawlers and any IP making more than 50 valid requests that day, following the economic logic in \citet{loughran2017edgar}; alternatives retain users up to 1,000 requests or rely only on SEC flags.

Five findings organize the paper. First, the statutory threshold is not a sharp adoption experiment in the reconstructed data. At the preferred bandwidth, the first-stage effect on filing-level XBRL use is only \FirstStage{} (SE \FirstStageSE{}). Below-threshold firms also adopted quickly. A sample selected on realized 2009 versus 2010 adoption produces a much stronger first stage, but that comparison conditions on a post-assignment outcome. Reporting both results turns a hidden design choice into a visible scientific result.

Second, estimated traffic changes differ across interfaces, although individual channel estimates are imprecise. The preferred intent-to-treat estimate for total unique-IP attention is \TotalITT{} log points, equivalent to approximately \TotalPct{} percent, while the primary-document estimate is \PrimaryITT{} and the XBRL/XML estimate is \XMLITT{}. The probability of observing at least one XML visitor rises by 0.138 (unadjusted $p=0.040$; within-family FDR $q=0.161$). Taken together, the channel estimates are suggestive of substitution toward structured assets even when direct total traffic does not rise; they are not decisive evidence of a causal reallocation.

Third, the modern filing architecture changes discontinuously even though classic human-facing complexity does not disappear. Inline XBRL rises from less than 3 percent of periodic filings in 2018 to approximately universal coverage among 10-Qs after 2022. Standardized SEC facts match \ModernFactCoverage{} percent of the modern census. The relevant disclosure object is therefore no longer a human document plus a peripheral data exhibit; it is a joint human--machine document connected to a near-real-time public API \citep{sec2018inline,secapis2025}.

Fourth, access becomes two-dimensional. In 2024, the rank correlation between document length and the composite machine-accessibility measure is \ModernWordMachineRho{}, not negative as a single processing-cost model would imply. The largest public-float quartile has both substantially higher human burden and substantially higher structured-data coverage than the smallest quartile. This is not evidence that large-firm investors are harmed or that AI systems actually consumed each filing. It is evidence that ``easy to read'' and ``easy for a machine to extract'' are empirically distinct attributes and that the machine-facing attribute is unequally distributed.

Fifth, randomized interface assignment causally changes grounded model outputs in the standardized numerical tasks, while the paired mechanism decomposition substantially narrows the interpretation of the initial bundled comparison. Better HTML localization closes most of the observed XBRL--HTML gap. When oracle text and oracle XBRL contain the same facts, the remaining XBRL advantage is only \FactorialSemanticEffect{} percentage points; replacing taxonomy-bearing XBRL with flat JSON changes accuracy by \FactorialTaxonomyEffect{} points. Accession-clustered and filing-bootstrap intervals rule out a one-percentage-point representation advantage under the matched-evidence design. The experimentally manipulated mechanism is therefore evidence addressability---retrieval and table reconstruction---rather than file syntax alone. This result replicates across six hosted model implementations and survives a counterfactual test in which original filing values appear only as decoys.

The paper makes three contributions. First, it refines the theory of cognitive load in financial disclosure by making processing cost explicitly reader-, task-, interface-, and stage-dependent. This reconciles classic human-oriented complexity measures with machine-mediated processing without relabeling machine accuracy as human cognition. Second, it provides causal evidence on one transformation-load mechanism: evidence addressability. The paired design separates localization and table reconstruction from representation syntax, taxonomy, and context length in standardized numerical tasks. Third, it develops transparent filing-level measures of human burden and machine accessibility and subjects the latter to initial, task-conditioned criterion validation. The historical reconstruction, modern census, counterfactual test, and auditable response archive supply distinct evidence for acquisition, transformation, and integration rather than forcing them into one outcome.

The results are scientifically useful partly because they constrain interpretation. We do not claim that a request equals attention, that an IP equals a person, or that model accuracy equals investor welfare. The historical estimates remain suggestive reduced forms; the modern census is descriptive; only assignment of information interfaces in the AI experiment is randomized. The experiment identifies interface effects across six hosted model implementations, one frozen response contract, five numerical task families, and the 2024 sample. That separation of causal, descriptive, and conceptual claims is essential.

The remainder describes the institutional transition and framework, presents the historical and modern evidence, reports the randomized AI experiment, and then develops implications and claim boundaries. Appendices document construction, scoring, robustness, and replication.

\section{Cognitive Load, Disclosure Interfaces, and Institutional Setting}

\subsection{What the mandate changes}

XBRL is a structured representation of financial-statement facts. A tagged fact has a concept, value, unit, reporting period, and entity context. This structure supports comparison and automated extraction. During the initial U.S. implementation, the structured layer was submitted as a set of exhibits alongside the conventional filing. The mandate did not replace the primary 10-K or 10-Q, nor did it make the prose intrinsically shorter or more readable.

The phase-in followed filer status and public float. The first cohort consisted of U.S.-GAAP large accelerated filers whose worldwide public float exceeded \$5 billion. The relevant compliance date was the fiscal-period end, not the filing submission date. This distinction matters for quarterly filings submitted several weeks after period end and corrects a timing error that can contaminate treatment assignment.

\begin{figure}[H]
\centering
\includegraphics[width=.94\textwidth]{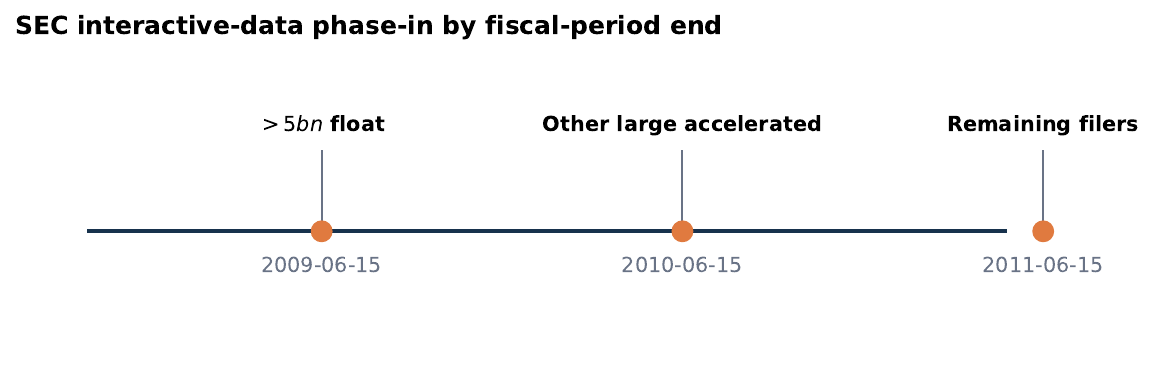}
\caption{SEC interactive-data phase-in. The dates refer to fiscal periods ending on or after the date shown. The empirical design focuses on the first transition and ends before the second cohort becomes subject to the mandate.}
\label{fig:timeline}
\end{figure}

The first-year requirement focused on block-tagged financial-statement information; more detailed footnote tagging followed. The initial transition thus combines at least three changes: availability of machine-readable statements, learning by filers and users, and evolving detail in the structured data. Our short window isolates the entry into XBRL more cleanly than a long post period, but it cannot separate all three components.

\subsection{Cognitive constraints, information acquisition, and disclosure processing}

Limited-attention models emphasize that investors allocate scarce processing capacity across signals \citep{hirshleifer2003limited,peng2006investor}. Empirical work documents distraction, media diffusion, and heterogeneous information demand around corporate news \citep{hirshleifer2009distraction,drake2012google,peress2014media}. Disclosure research shows that length, readability, and presentation affect use by different investor classes \citep{miller2010disclosure,lawrence2013individual,loughran2014measuring}. The modern processing-cost perspective treats acquisition, transformation, and integration as separate tasks \citep{blankespoor2020processing}. XBRL targets the transformation task especially directly.

Prior XBRL work generally studies downstream outcomes. \citet{dong2016information} link adoption to firm-specific information acquisition through return synchronicity and price delay. \citet{cheng2021investment} study investment efficiency. Other research documents data-quality problems and firm responses to detailed tagging \citep{debreceny2011xbrl}. These studies motivate an information-cost channel but do not make direct server traffic a complete measure of effective use.

EDGAR logs allow a closer view of acquisition. \citet{drake2015edgar} study determinants and consequences of EDGAR information acquisition, and \citet{loughran2017edgar} show why automated requests must be treated carefully. \citet{cong2018accessed} directly compare XBRL and conventional file accesses for small companies after the broad phase-in. Our contribution relative to this work is not the observation that XBRL files receive requests. It is the combination of transition timing, predetermined assignment, document-level channel decomposition, cross-day deduplication, and estimand sensitivity to automated users.

\subsection{The measurement problem}

An EDGAR row identifies an anonymized IP, date, time, accession, requested extension, response code, and several flags. It does not identify a beneficial owner, a decision, or a downstream audience. Three examples illustrate the ambiguity.

First, one analyst may access the same filing from several IPs. Second, one data vendor may download a structured file once and serve hundreds of clients. Third, a high-volume request can be either a low-value crawler or a high-value information intermediary. A filter that removes the latter makes the outcome more human-like but less representative of the technology's intended machine use. For this reason, our main tables label every traffic definition explicitly.

\subsection{From XBRL to an AI-mediated disclosure supply chain}

The original XBRL rule made standardized facts available alongside the conventional filing. The 2018 Inline XBRL rule subsequently embedded machine-readable tags inside the human-readable document, with a phased compliance schedule through 2021 \citep{sec2018inline}. The SEC now distributes submissions metadata and standardized Company Facts through unauthenticated JSON APIs and nightly bulk archives \citep{secapis2025}. These changes reduce the fixed cost of building systems that retrieve, normalize, and compare filings. They also make one acquisition event capable of supporting many downstream interfaces.

Large language models extend this transformation margin beyond numerical extraction. They can summarize, classify, retrieve, and synthesize narrative content, although accuracy and provenance remain task dependent. \citet{kim2023bloated} show that model-generated summaries can compress disclosure ``bloat'' and retain information related to market reactions. Finance-specific language models and benchmarks document rapid improvement in domain tasks \citep{wu2023bloomberggpt,yang2023fingpt}. These studies establish technological feasibility; they do not reveal how many investors use the resulting outputs or whether access is equally distributed. Our framework therefore treats AI as a downstream transformation technology and structured disclosure as observable infrastructure, not as an observed treatment assigned to every filing.

Recent work makes the representation problem more specific. XBRL-Agent shows that retrieval and explicit calculation tools improve question answering over structured reports, while FinTagging finds that extracting financial entities is substantially easier than mapping them to fine-grained taxonomy concepts \citep{han2024xbrlagent,wang2026fintagging}. Those studies develop systems and benchmarks; our factorial design instead holds the question and underlying filing facts fixed while independently varying localization, representation, and irrelevant context. This distinction matters because a large XBRL--HTML difference can otherwise be generated by the retriever rather than by taxonomy semantics. We also address two threats that become acute for historical financial questions: model outputs are not perfectly reproducible across deployments, and public filing values may have entered pretraining data \citep{wang2025reproducibility,lopezlira2025memorization}. Our multi-provider replication estimates interface effects separately by model family, and a counterfactual-evidence test replaces target values while retaining original values as plausible decoys. Finally, emerging evidence links firms' tagging choices to AI-equipped investor exposure \citep{engel2026aiequipped}. We study the complementary demand-side question: which properties of a filing interface actually improve a model's grounded use of disclosure?

\subsection{Readability, processing cost, and the identity of the reader}

Annual-report readability predicts trading, analyst effort, and market outcomes in human-oriented settings \citep{li2008annual,miller2010disclosure,lehavy2011effect}. Experiments further show that processing fluency and benchmark consistency shape investor judgments \citep{rennekamp2012processing,tan2015readability}. Yet readability is not a universal physical property of a file. Fog-style language scores have known limitations in financial text, and file size can outperform linguistic formulas as an empirical proxy \citep{loughran2014measuring}. More recent measures explicitly model plain-English attributes or disclosure bloat \citep{bonsall2017plain,kim2023bloated}. Textual-change, cross-country, and risk-factor studies likewise show that the appropriate representation depends on the economic task \citep{brown2011mda,lang2015textual,campbell2014risk}. Each measure is useful for a declared reader and task; none is guaranteed to rank processing cost identically for a structured-data parser, a retrieval system, and an unaided retail investor.

The most closely related evidence is \citet{call2024human}, who use the Inline XBRL mandate to study whether improved machine readability changes firms' human-readable disclosure choices and retail-investor outcomes. Their question concerns a real effect of machine-readability regulation on human readability. Our question concerns measurement invariance: after the interface changes, can a human-readable complexity measure stand in for effective access, and can a source-server request stand in for attention? The papers are complementary, but the distinction determines our empirical objects and conservative interpretation.

\subsection{Attention and information fairness}

Limited-attention theory treats processing capacity as scarce \citep{hirshleifer2003limited,peng2006investor}. Empirical work uses announcement timing, salient stocks, internet searches, and disclosure acquisition to observe allocation of that capacity \citep{dellavigna2009investor,barber2008all,da2011search,drake2015edgar}. Machine intermediation does not eliminate scarcity; it changes where scarcity enters. The bottleneck may shift from finding and reading a document to selecting a model, verifying an answer, integrating data, or paying for an intermediary. Consequently, an unchanged traffic proxy can load on different latent tasks across technological regimes.

Information fairness likewise has more than one margin. The business press and public dissemination can broaden reach and reduce selective-access advantages \citep{bushee2010businesspress,blankespoor2014initial}, while information acquisition and processing technologies can create heterogeneous effective access even when raw documents are public. In a rational-expectations setting, costly information and differential processing capacity affect who becomes informed and how prices aggregate signals \citep{grossman1980impossibility,veldKamp2006media,easley2016differential}. Data abundance can improve price informativeness while changing private incentives to acquire and process signals \citep{dugast2018data,farboodi2020data}. Our evidence speaks to an infrastructure prerequisite for this channel: machine-ready content is more abundant for some issuers. It does not identify portfolio returns, welfare, or a legal violation of fair disclosure.

\begin{figure}[H]
\centering
\includegraphics[width=.98\textwidth]{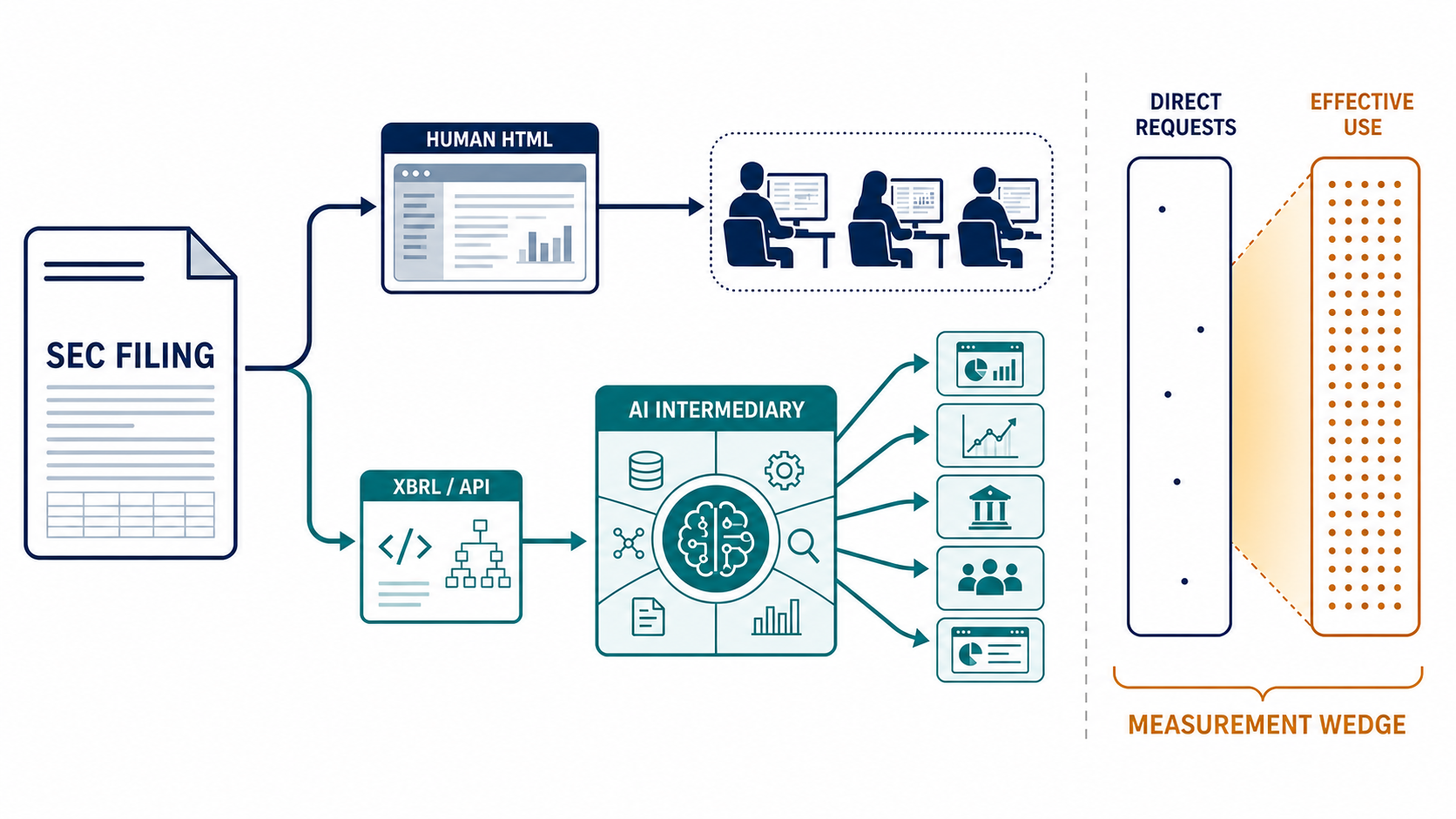}
\caption{Disclosure supply chain and the observability wedge. The upper path depicts direct human acquisition of the HTML filing. The lower path depicts machine-readable content transformed by an intermediary and redistributed through multiple products. The right-hand contrast is conceptual: direct requests are observed at EDGAR, whereas downstream effective use is not. The illustration is not an empirical estimate.}
\label{fig:aiwedge}
\end{figure}

\section{A Reader--Task--Interface Framework of Cognitive Load}

The framework makes the processing channel explicit. Cognitive load is not treated as a psychological score recovered from a document alone. It is the task-contingent cost of acquiring, transforming, and integrating evidence under a reader's available interface. The same filing can therefore generate different loads for an unaided investor, a structured-data parser, and an AI intermediary. The claim is deliberately modest: the present data identify interface-dependent processing outcomes, not individual working-memory capacity or investor welfare.

Let $r$ index a reader or reading technology, $t$ a task, and $k$ an interface. We decompose effective processing load as
\begin{equation}
L_{rtk}=L^{A}_{rtk}+L^{T}_{rtk}+L^{I}_{rtk},
\label{eq:loaddecomp}
\end{equation}
where $L^{A}$ is acquisition load, $L^{T}$ is transformation load required to locate and reconstruct task-relevant evidence, and $L^{I}$ is integration load required to use the evidence consistently and with adequate support. Traditional readability measures primarily describe inputs to $L^{T}$ and $L^{I}$ for a human reader. Source-server requests are partial observations of $L^{A}$. The AI experiment below operationalizes the transformation and integration components through evidence coverage, grounded correctness, citation validity, abstention, and schema compliance.

\noindent\textbf{Definition (interface-local cognitive load).} A complexity measure is interface-local if it ranks two filings differently after either the reader, task, or representation changes. Measurement invariance therefore requires an explicit declaration of $(r,t,k)$; it cannot be assumed from a stable file-level score.

\subsection{Users, tasks, and interfaces}

Consider a filing containing economic content $z$ and two interfaces, $k\in\{H,X\}$: the primary human-readable document $H$ and the structured XBRL layer $X$. User $i$ performs task $t\in\{R,E\}$, where $R$ is contextual reading and $E$ is numerical extraction. The gross value of completing the task is $v_{it}(z)$. Processing interface $k$ entails cost
\begin{equation}
c_{itk}=a_{itk}+b_{tk}C+\eta_{ik},
\end{equation}
where $C$ is disclosure complexity, $a_{itk}$ is an interface-specific fixed cost, and $\eta_{ik}$ captures user capability and software. The user selects the lowest-cost feasible interface and acquires information when
\begin{equation}
v_{it}(z)-\min_{k\in\mathcal K_t}c_{itk}\geq 0.
\end{equation}
Before XBRL, $X$ is unavailable or has prohibitive cost for standardized extraction. The mandate reduces $a_{iEX}$ for users with suitable tools. It need not reduce $c_{iRH}$, because the narrative filing remains.

\noindent\textbf{Implication 1 (Interface substitution).} If the mandate weakly reduces $c_{iEX}$ and leaves all other costs unchanged, the share of extraction tasks routed through $X$ weakly increases. The number of primary-document requests need not increase and can decrease when $H$ and $X$ are substitutes for extraction.

The result follows directly from revealed preference: lowering one option's cost cannot reduce its choice set, but it can draw tasks from another interface. This prediction is about composition, not total attention.

\subsection{Intermediaries and observable traffic}

Let a direct user generate one observable request when acquiring from EDGAR. An intermediary $m$ generates $q_m$ EDGAR requests and serves $s_m\geq q_m$ downstream task instances. Effective information acquisition is
\begin{equation}
A^{\ast}=A^{D}+\sum_m s_m,
\end{equation}
whereas observed EDGAR traffic is
\begin{equation}
A^{E}=A^{D}+\sum_m q_m.
\end{equation}
XBRL can raise the distribution multiplier $s_m/q_m$ by allowing one structured download to populate databases, screens, and models.

\noindent\textbf{Implication 2 (Ambiguous observable traffic).} Even if the mandate weakly increases effective acquisition $A^{\ast}$, its effect on observable unique-IP traffic $A^{E}$ is ambiguous. The effect is negative when substitution and intermediary consolidation exceed new direct entry.

This implication prevents a common inferential leap. A fall in EDGAR IPs does not prove less information use; a rise does not prove better comprehension. The channel composition and the automated-user definition provide additional evidence, but downstream use remains unobserved.

\subsection{Predictions}

The framework yields four empirically disciplined predictions.

\begin{prediction}[Machine-readable share]
Statutory assignment increases the probability and share of filing visitors requesting XBRL/XML assets, conditional on a first stage into observed XBRL.
\end{prediction}

\begin{prediction}[Primary-document substitution]
Assignment can reduce primary-document traffic when numerical extraction migrates from HTML to XBRL.
\end{prediction}

\begin{prediction}[Robot-rule sensitivity]
The measured total effect differs across robot definitions because high-volume automated users are part of the machine-readable disclosure channel.
\end{prediction}

\begin{prediction}[Complexity]
Interface substitution may be stronger for numerically or structurally complex filings, but text length alone need not amplify the effect because XBRL does not structure all narrative content.
\end{prediction}

These are predictions about observable acquisition. They do not assert price efficiency, errors, welfare, or investor sophistication without corresponding data.

\subsection{Construct validity under technological change}

Three system properties must be separated. \emph{Availability} means that information publicly exists; \emph{addressability} means that a system can locate, disambiguate, and reconstruct the task-relevant evidence; \emph{grounded usability} means that an agent can correctly use that evidence and return auditable support. They form neither identities nor automatic implications:
\begin{equation}
\text{Availability}\ \not\Rightarrow\ \text{Addressability}\ \not\Rightarrow\ \text{Grounded usability}.
\label{eq:three_layers}
\end{equation}
The distinction is operational rather than semantic: the experiment separately records required-fact coverage, numerical correctness, evidence validity, abstention, and schema compliance.

Let $L_{jt}$ be the latent effective attention devoted to filing $j$ at time $t$, and let $Q_{jt}$ be source-server requests. In a direct-reading regime, a simple measurement equation is
\begin{equation}
Q_{jt}=\alpha_t+\lambda_t L_{jt}+u_{jt}, \qquad \lambda_t>0.
\end{equation}
If each acquisition serves one task, comparisons in $Q$ can be informative about comparisons in $L$. With intermediaries, let $M_{jt}$ be machine acquisitions and $s_{mjt}$ the unobserved downstream service multiplier. Effective attention becomes
\begin{equation}
L_{jt}=H_{jt}+\sum_m s_{mjt}M_{mjt},
\end{equation}
while the server observes only $Q_{jt}=H_{jt}+\sum_m M_{mjt}$. Unless the distribution of $s_{mjt}$ is constant, the loading $\lambda_t$ is technology dependent. This is a failure of measurement invariance, not merely classical noise. It can reverse comparisons across filings or time.

The same logic applies to processing cost. Define $B^H_j$ as human processing burden and $A^M_j$ as machine accessibility. A one-dimensional disclosure score assumes that a monotone transformation $g$ maps the observed document metric $R_j$ to both constructs. The joint-interface regime instead permits
\begin{equation}
B^H_j=g_H(W_j,S_j,V_j), \qquad
A^M_j=g_M(T_j,F_j,C_j,P_j),
\end{equation}
where $W$ is length, $S$ syntax, $V$ vocabulary, $T$ inline tags, $F$ standardized fact observations, $C$ concept coverage, and $P$ parser accessibility. No theoretical restriction requires $B^H$ and $A^M$ to be negatively correlated. A long filing can be difficult for a person and rich in machine-readable facts.

\begin{proposition}[Metric non-invariance]
If either the intermediary service multiplier varies across filings or the ordering of $B^H$ differs from the ordering of $-A^M$, then a scalar request or readability measure cannot preserve the ranking of effective access across both direct-reading and machine-intermediated regimes.
\end{proposition}

The proposition motivates two empirical diagnostics. First, historical traffic effects are decomposed by interface and robot definition. Second, modern filings are ranked separately on transparent human-burden and machine-access primitives. We treat rank concordance as evidence about construct overlap, not as a test that one metric is intrinsically superior.

\subsection{A two-interface definition of information fairness}

Let investor $i$'s effective access to filing $j$ be
\begin{equation}
\mathcal A_{ij}=\max\{a^H_{ij},a^M_{ij}\}-p_{ij},
\end{equation}
where $a^H$ depends on literacy and time, $a^M$ depends on tools and data coverage, and $p$ is the monetary and verification cost of the chosen channel. Document-level equality guarantees that investors face the same raw filing, not that they face the same $a^M$ or $p$. Conversely, broad public APIs can democratize structured access when tool costs fall. The fairness effect is therefore theoretically ambiguous.

We use the term \emph{infrastructure fairness} narrowly: whether machine-readable coverage differs systematically across public-float groups under a formally universal disclosure regime. This is a necessary input into broader information fairness, not a sufficient statistic for it. Demonstrating unequal tag or fact coverage cannot establish unequal beliefs, trades, or welfare; demonstrating equal coverage cannot establish equal model quality or verification ability.

\section{Data Reconstruction}

\subsection{Filing universe and sample flow}

We begin with the SEC bulk submissions archive and enumerate all non-amended Forms 10-K and 10-Q filed from 2008 through 2014. The universe has 238,566 filings from 15,899 CIKs. For each filing we retain accession number, form, filing date, fiscal-period end, acceptance timestamp, primary-document filename, and the SEC XBRL indicator.

The causal window uses fiscal periods ending from June 15, 2008 through June 14, 2010. Ending the window before the second phase prevents the below-threshold comparison group from becoming statutorily treated. Filings may arrive through August 31, 2010 to accommodate reporting lags. We require at least two pre- and two post-period filings.

\begin{figure}[H]
\centering
\includegraphics[width=.98\textwidth]{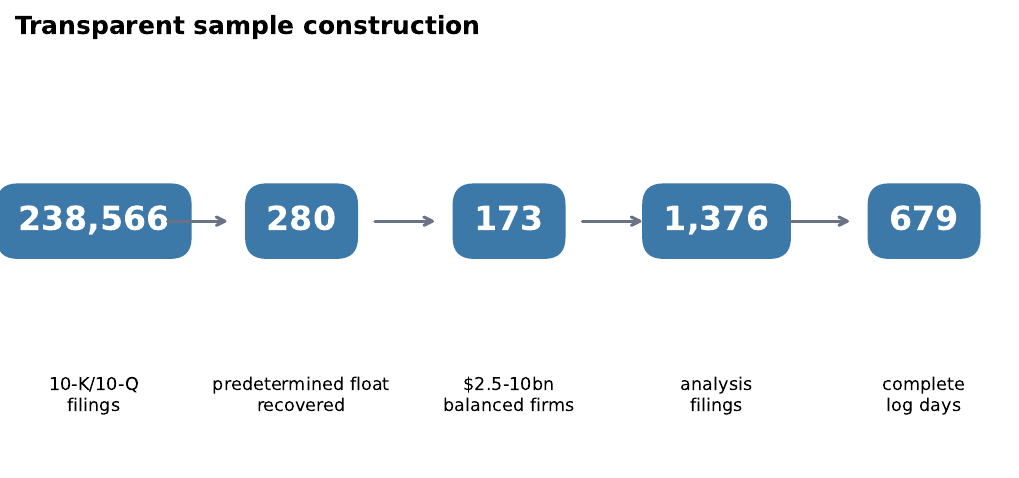}
\caption{Sample construction. Public float is recovered independently of realized adoption. All analysis filings have a primary document and a complete eight-day EDGAR-log window.}
\label{fig:flow}
\end{figure}

\subsection{Predetermined public float}

The assignment variable is the SEC's standardized \texttt{dei:EntityPublicFloat} fact with a measurement date from January 1, 2008 through June 14, 2009. We use the latest eligible measurement date, the modal value when a fact is repeated, and the earliest filing as a tie-breaker. Values outside \$0.5--20 billion are rejected before applying the analysis bandwidth. The main candidate range is \$2.5--10 billion.

Most historical facts are recovered retrospectively from later structured submissions. Their measurement dates predate the mandate, but their structured representation was filed later. This is superior to selecting on realized adoption yet inferior to a contemporaneously archived cover-page database. We therefore disclose the recovery filing date, report sensitivity, and avoid claiming a textbook regression discontinuity.

\subsection{EDGAR traffic}

We download each required daily log from the SEC's public dataset \citep{seclogs}. A valid row has a response code below 300, is not marked as a crawler, is not an index-page request, and contains an IP and accession. Each row is matched exactly to an accession. On event day zero, only requests at or after the filing acceptance time count.

The preferred screen also excludes an IP if it makes more than 50 valid EDGAR requests anywhere that day. We report two alternatives: excluding only IPs above 1,000 requests and using the SEC crawler flag alone. Counts are constructed at two levels. ``IP-days'' deduplicate within a date and sum across dates. ``Unique IPs'' take the union across the complete days 0--7 window. The latter is primary.

Requests are decomposed into: (i) any asset associated with the accession, (ii) the exact primary document, and (iii) files ending in \texttt{.xml} or \texttt{.xsd}. The third category is a transparent proxy for machine-readable use. It omits some supporting files and does not identify the purpose of a request.

\subsection{Filing text and complexity}

Every primary document is downloaded from the SEC archive. Scripts and styles are removed, whitespace is normalized, and transparent features are computed: bytes, characters, words, sentences, words per sentence, mean word length, long-word share, numeric tokens per thousand words, tables, links, and inline-XBRL tags. Heterogeneity uses firm-level pre-policy means, so post-treatment document changes do not define subgroups.

We deliberately avoid labeling these variables ``cognitive load.'' Word count measures information quantity; sentence length measures syntax imperfectly; table and numeric density capture structured content. They are observable document characteristics rather than direct psychological measures.

\subsection{Descriptive evidence}

\begin{figure}[H]
\centering
\includegraphics[width=.82\textwidth]{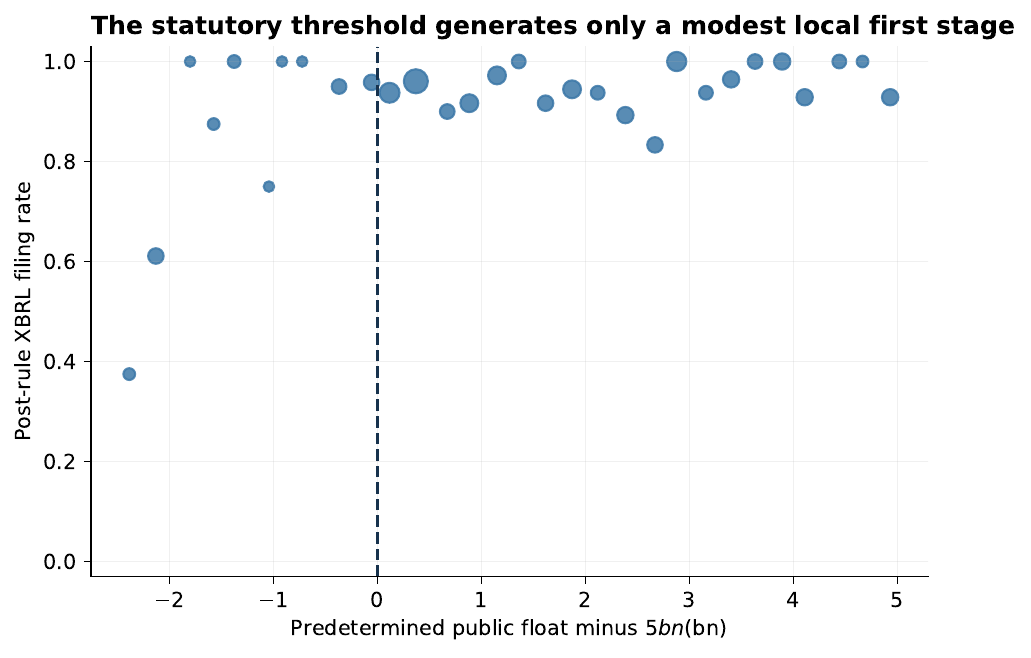}
\caption{Predetermined public float and post-rule XBRL use. Marker size is proportional to the number of firms in each public-float bin. The plot is descriptive and does not condition on the realized adoption cohort.}
\label{fig:firststage}
\end{figure}

The final independent sample has \SampleFirms{} firms and \SampleFilings{} filings. Figure \ref{fig:firststage} plots post-period XBRL rates against predetermined public float. The local discontinuity is visibly smaller than the nominal phase-in might suggest.

\begin{table}[H]
\centering
\caption{Descriptive Cells}
\label{tab:descriptive}
\begin{threeparttable}
\begin{tabular}{llrrrrr}
\toprule
Assignment & Period & Filings & Firms & XBRL rate & Mean IPs & Mean XML IPs \\
\midrule
Below $5$bn & Pre & 116 & 29 & 0.009 & 51.9 & 0.4 \\
Below $5$bn & Post & 113 & 29 & 0.832 & 72.3 & 14.8 \\
Above $5$bn & Pre & 571 & 144 & 0.016 & 55.3 & 0.5 \\
Above $5$bn & Post & 576 & 144 & 0.948 & 70.4 & 12.7 \\
\bottomrule
\end{tabular}
\begin{tablenotes}[flushleft]\footnotesize
\item Notes: IP measures use unique IPs deduplicated across filing days 0--7 after the daily 50-request screen. Cells are unadjusted.
\end{tablenotes}
\end{threeparttable}
\end{table}

\section{Empirical Design: Acquisition and Processing Tests}

\subsection{Difference around the threshold}

Let $Z_i$ equal one when predetermined public float exceeds \$5 billion, $Post_{it}$ indicate a fiscal period ending on or after June 15, 2009, and $R_i$ denote public float minus the threshold in billions. For filing $f$ of firm $i$, we estimate
\begin{equation}
y_{if}=\alpha_i+\lambda_{m(f)}+\phi_{form(f)}+\beta(Z_i\times Post_{if})
+\rho(R_i\times Post_{if})+\varepsilon_{if}.
\label{eq:main}
\end{equation}
Firm fixed effects absorb time-invariant level differences, filing-month effects absorb aggregate traffic and crisis conditions, and form effects distinguish 10-K from 10-Q. Standard errors cluster by firm. The coefficient $\beta$ is an intent-to-treat difference associated with threshold assignment.

The preferred bandwidth is \$2.5 billion on either side where support permits, with \$1 billion and the full asymmetric available range reported as sensitivity. Because the recovered sample contains fewer firms below the threshold and the first stage is fuzzy, we do not present a bias-corrected sharp-RD estimate or a treatment-on-the-treated ratio as the headline.

\subsection{First stage and channel outcomes}

The first stage replaces $y_{if}$ with the filing-level SEC XBRL indicator. Channel outcomes include log unique IPs to any filing asset, the primary document, and XML/XSD assets; an indicator for any XML visitor; and XML and primary-document visitor shares. Shares use total unique filing IPs as the denominator and are set to zero when no eligible visit occurs.

\subsection{Dynamic and diagnostic specifications}

The event study replaces $Z_i\times Post_{if}$ with interactions between $Z_i$ and fiscal quarters relative to 2009Q2, omitting the immediately preceding quarter. End bins collect observations at least four quarters before or three quarters after. Dynamic coefficients are descriptive diagnostics because the design has one assignment threshold and relatively few below-threshold firms.

We also estimate day-specific specifications based on Equation \ref{eq:main}, two placebo policy dates using pre-period observations only, local covariate-continuity regressions, and heterogeneity by predetermined filing characteristics. False-discovery-rate adjusted $q$-values are reported within outcome families. We emphasize effect sizes and confidence intervals rather than binary significance labels.

\subsection{Identification assumptions and threats}

The design requires that firms just above and below the recovered threshold would have followed comparable changes absent the first phase. Three threats deserve emphasis. First, public float can be noisy and recovered retrospectively. Second, some below-threshold firms adopted voluntarily or appear to have been subject to other eligibility details. Third, the financial crisis overlaps the pre-period. Firm and filing-month effects address levels and common shocks, not differential shocks correlated with the threshold.

The design does not require that total traffic rise. It requires only that threshold assignment changes the probability of structured filing availability for the channel interpretation. A weak first stage makes reduced-form evidence less precise and prevents a stable Wald ratio.

Selecting firms by whether they actually adopted in 2009 or 2010 raises the apparent first stage from 0.113 to 0.329 and makes the reduced form look more precise. Because realized adoption is downstream of assignment, the predetermined sample is primary; Figure \ref{fig:selection} reports the comparison in the appendix.

\section{Results: Acquisition, Infrastructure, and Machine Processing}

\subsection{The first stage is informative but modest}

At the preferred bandwidth, threshold assignment raises filing-level XBRL use by \FirstStage{} (SE \FirstStageSE{}). This first stage is statistically and economically smaller than the phase-in description alone suggests. The reason is visible in the raw cells: many below-threshold filings also have XBRL in the post period.

This finding changes the interpretation of the setting. Comparing firms whose \emph{observed} first adoption occurs in 2009 with those adopting in 2010 mechanically sharpens the first stage, but adoption timing is an outcome. The independent sample instead treats imperfect compliance as evidence. The mandate remains informative, but it is not a clean on/off switch.

\subsection{Acquisition load and interface composition}

\begin{figure}[H]
\centering
\includegraphics[width=.82\textwidth]{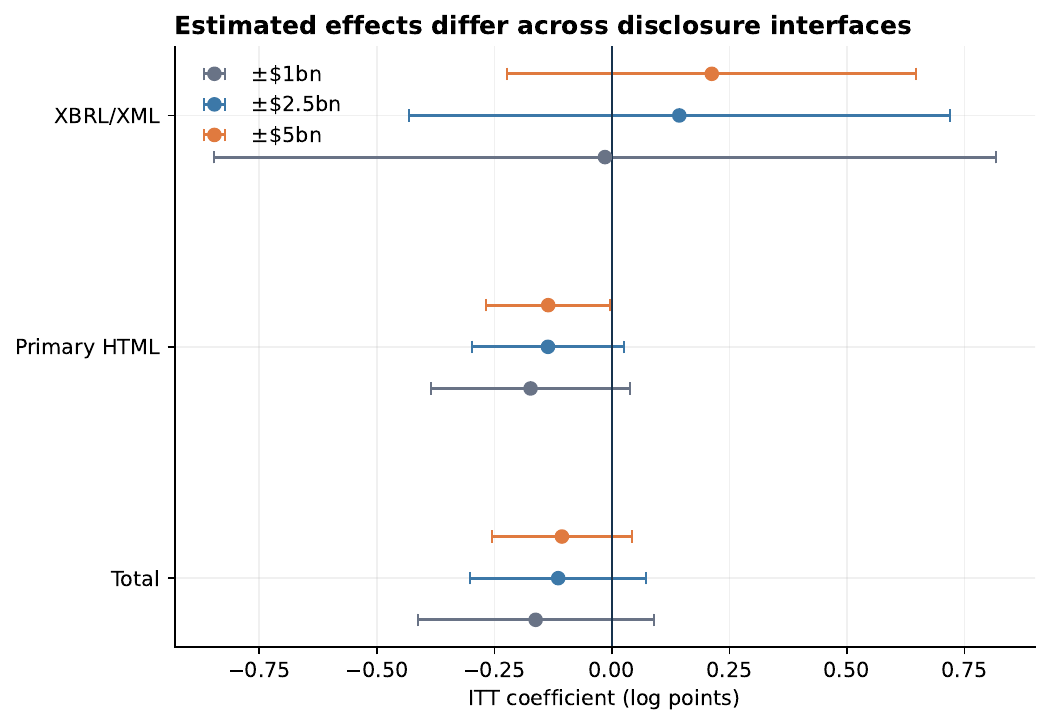}
\caption{Channel decomposition across bandwidths. Points are intent-to-treat coefficients; bars are 95 percent firm-clustered confidence intervals. The widest range reflects the asymmetric support available above the threshold.}
\label{fig:channels}
\end{figure}

The preferred total-acquisition estimate is \TotalITT{} log points (SE \TotalITTSE{}), or approximately \TotalPct{} percent. Primary-document acquisition changes by \PrimaryITT{} (SE \PrimaryITTSE{}), whereas XBRL/XML acquisition changes by \XMLITT{} (SE \XMLITTSE{}). Figure \ref{fig:channels} shows all three estimates across bandwidths.

The joint pattern is more informative than the total coefficient. Primary-document visits need not increase when structured extraction becomes available. XML use can expand sharply from a low base. Depending on which users are retained, total unique IPs can be flat or lower even if structured information reaches additional downstream users.

\begin{table}[H]
\centering
\caption{Intent-to-Treat Effects across Bandwidths}
\label{tab:main}
\begin{threeparttable}
\begin{tabular}{lccc}
\toprule
Outcome & $\pm$1bn & $\pm$2.5bn & Available range \\
\midrule
Total unique IPs & -0.162\;(0.128) & -0.114\;(0.096) & -0.106\;(0.076) \\
Primary-document IPs & -0.173\;(0.108) & -0.136\;(0.082) & -0.135\;(0.067) \\
XBRL/XML IPs & -0.015\;(0.424) & 0.144\;(0.294) & 0.213\;(0.222) \\
Observed XBRL & 0.090\;(0.073) & 0.113\;(0.065) & 0.140\;(0.056) \\
Any XML visitor & 0.117\;(0.072) & 0.138\;(0.067) & 0.147\;(0.057) \\
\bottomrule
\end{tabular}
\begin{tablenotes}[flushleft]\footnotesize
\item Notes: Filing-level regressions include firm, filing-month, and form fixed effects plus a post-period running-variable slope. Firm-clustered standard errors are in parentheses. Traffic outcomes are log(1 + unique IPs) over days 0--7 under the daily 50-request screen. The available range is asymmetric because support below the threshold ends at -$2.5$bn.
\end{tablenotes}
\end{threeparttable}
\end{table}

The results therefore support a limited interface-substitution conclusion about the acquisition stage of Equation \ref{eq:loaddecomp}. They do not establish that human comprehension improved, that investors made fewer errors, or that prices incorporated information faster. Those outcomes require behavioral or market data not present in the public reconstruction.

Dynamic coefficients, raw trends, and day-specific estimates are reported in Appendix Figures \ref{fig:eventstudy}--\ref{fig:days}. They reveal secular traffic variation and imprecise event-day timing; none upgrades the historical module beyond a diagnostic.

\section{Disclosure Complexity across Human and Machine Interfaces, 2018--2025}

\subsection{Why a second empirical module is necessary}

The 2009 transition offers exact source-server traffic and a policy assignment variable, but it predates Inline XBRL, public Company Facts APIs, and general-purpose language models. Extrapolating its traffic coefficients directly to an AI-mediated market would be scientifically indefensible. The modern module asks a different question with different identification: what filing infrastructure now exists, and do traditional human-facing complexity measures span the same empirical variation as machine accessibility? The answer is descriptive but central to external validity.

We use two official SEC bulk archives downloaded as immutable ZIP files. The submissions archive contains public filing histories, primary-document filenames, and XBRL/Inline-XBRL indicators. The Company Facts archive contains standardized facts extracted by the SEC, including accession, concept, unit, period, and filing metadata. Both are nightly bulk counterparts of public, unauthenticated APIs \citep{secapis2025}. We retain non-amended Forms 10-K and 10-Q filed from January 1, 2018 through December 31, 2025 and deduplicate on CIK--accession.

The resulting census contains \ModernCensusFilings{} filings by \ModernCensusFirms{} CIKs. For every accession we count standardized fact observations, distinct standardized concepts, and represented taxonomy namespaces. We match the SEC's 
\texttt{EntityPublicFloat} when attached to the same annual filing. Because quarterly filings rarely carry public float, size-stratified document analysis is restricted to 10-Ks. A standardized-fact count is an observable measure of public machine-ready coverage. It is not a measure of semantic accuracy, extension quality, model reasoning, or realized AI use.

\subsection{A fixed, balanced document sample}

Parsing every modern primary document would add substantial bandwidth and computation without improving the policy-adoption census. We therefore declare a separate diagnostic sample before extracting text. Eligible observations are 10-Ks with positive same-accession public float in 2018, 2020, 2022, or 2024. Within each year, issuers are assigned to public-float quartiles. A deterministic random seed selects 25 reports from each year--quartile cell, yielding \ModernSampleFilings{} filings from \ModernSampleFirms{} issuers. The 16 cells are exactly balanced.

This design is not representative weighting for the population of issuers; it deliberately supplies common support across scale and transition year. Results are reported both as transparent medians and as regressions with filing-year and two-digit SIC fixed effects. The selection algorithm, seed, accession list, raw documents, and failures file are included in the replication package. All 400 primary documents downloaded successfully.

For each document we compute bytes, visible words, heuristic sentences, words per sentence, mean word length, long-word share, numeric-token density, table count, link count, and Inline XBRL tag count. BeautifulSoup with the lxml parser removes scripts and styles before text extraction. Machine coverage adds standardized fact observations and distinct concepts from Company Facts. No proprietary model output enters the primary indices.

\subsection{Declared indices and primitive measures}

The human-burden index is the equal-weight mean of within-year standardized log word count, words per sentence, and mean word length. The machine-accessibility index is the equal-weight mean of within-year standardized log fact observations, log distinct concepts, and log Inline XBRL tags. Equal weights are declared for interpretability, and every primitive remains visible in tables and figures. The interface-divergence index is
\begin{equation}
D_j=B^H_j+A^M_j.
\end{equation}
It is high when a filing is relatively burdensome for unaided human reading and relatively accessible to structured-data tools. Because both components are standardized within filing year, $D$ compares issuers within the contemporary technological regime rather than mechanically encoding the phase-in.

The index is a diagnostic coordinate, not a cardinal welfare measure. Words per sentence is only a syntax heuristic; mean word length can reflect necessary technical vocabulary; tag counts can include duplicated contexts; Company Facts omits some custom-tag information; and more facts are not automatically better facts. We therefore make conclusions only when index patterns are also visible in the primitive variables.

\begin{figure}[H]
\centering
\includegraphics[width=.84\textwidth]{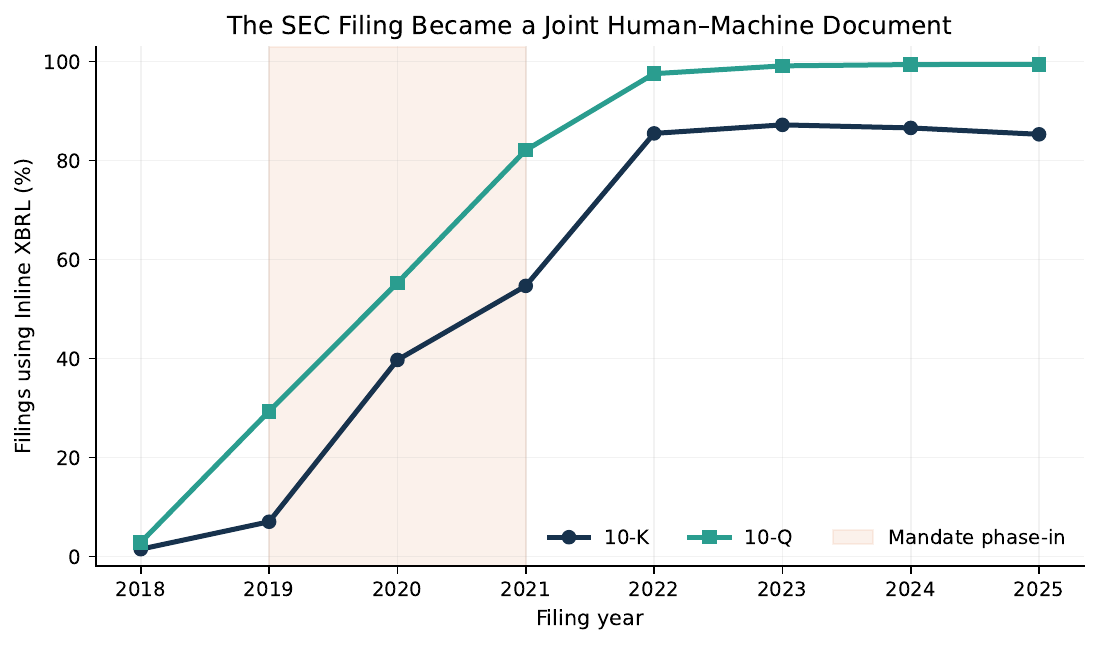}
\caption{Inline XBRL adoption in the modern SEC census. Shares use all \ModernCensusFilings{} non-amended 10-K and 10-Q filings in the 2018--2025 bulk submissions archive. The shaded region marks the operating-company phase-in. This is a census description, not a causal estimate.}
\label{fig:inlineadoption}
\end{figure}

Figure \ref{fig:inlineadoption} shows the architectural transition. In 2018 fewer than three percent of periodic filings use Inline XBRL. Adoption rises first among the phased-in groups and is nearly universal for 10-Qs after 2022. The 10-K share remains lower because the filing universe includes entities and reporting situations not uniformly subject to the operating-company requirement. The central implication does not depend on exactly 100 percent coverage: the modal modern filing is a joint human--machine document.

\begin{table}[H]
\centering
\caption{Modern SEC Filing Infrastructure}
\label{tab:moderncensus}
\begin{threeparttable}
\begin{tabular}{lrrrrrr}
\toprule
Year & Form & Filings & Issuers & Inline (\%) & Fact coverage (\%) & Median concepts \\
\midrule
2018 & 10-K & 7,092 & 6,907 & 1.5 & 85.3 & 165 \\
2018 & 10-Q & 19,216 & 6,549 & 2.8 & 96.0 & 112 \\
2019 & 10-K & 6,923 & 6,777 & 7.0 & 85.5 & 165 \\
2019 & 10-Q & 18,485 & 6,327 & 29.3 & 96.8 & 129 \\
2020 & 10-K & 6,769 & 6,615 & 39.7 & 84.9 & 176 \\
2020 & 10-Q & 17,979 & 6,235 & 55.3 & 97.2 & 125 \\
2021 & 10-K & 7,181 & 6,984 & 54.6 & 84.0 & 164 \\
2021 & 10-Q & 20,393 & 7,073 & 82.1 & 97.2 & 119 \\
2022 & 10-K & 7,792 & 7,648 & 85.4 & 85.8 & 162 \\
2022 & 10-Q & 20,758 & 7,104 & 97.5 & 97.7 & 120 \\
2023 & 10-K & 7,366 & 7,243 & 87.2 & 87.1 & 171 \\
2023 & 10-Q & 19,103 & 6,608 & 99.1 & 99.1 & 123 \\
2024 & 10-K & 6,878 & 6,768 & 86.6 & 86.3 & 173 \\
2024 & 10-Q & 17,770 & 6,146 & 99.4 & 99.1 & 124 \\
2025 & 10-K & 6,690 & 6,558 & 85.3 & 85.0 & 173 \\
2025 & 10-Q & 17,289 & 5,974 & 99.4 & 99.2 & 125 \\
\bottomrule
\end{tabular}
\begin{tablenotes}[flushleft]\footnotesize
\item Notes: Census of non-amended Forms 10-K and 10-Q in SEC bulk submissions, 2018--2025. Fact coverage equals the share matched to at least one observation in SEC Company Facts. Issuer counts are distinct CIKs within row.
\end{tablenotes}
\end{threeparttable}
\end{table}

\section{Is Disclosure Complexity Reader-Invariant?}

\subsection{Human burden and machine access occupy two dimensions}

\begin{figure}[H]
\centering
\includegraphics[width=.98\textwidth]{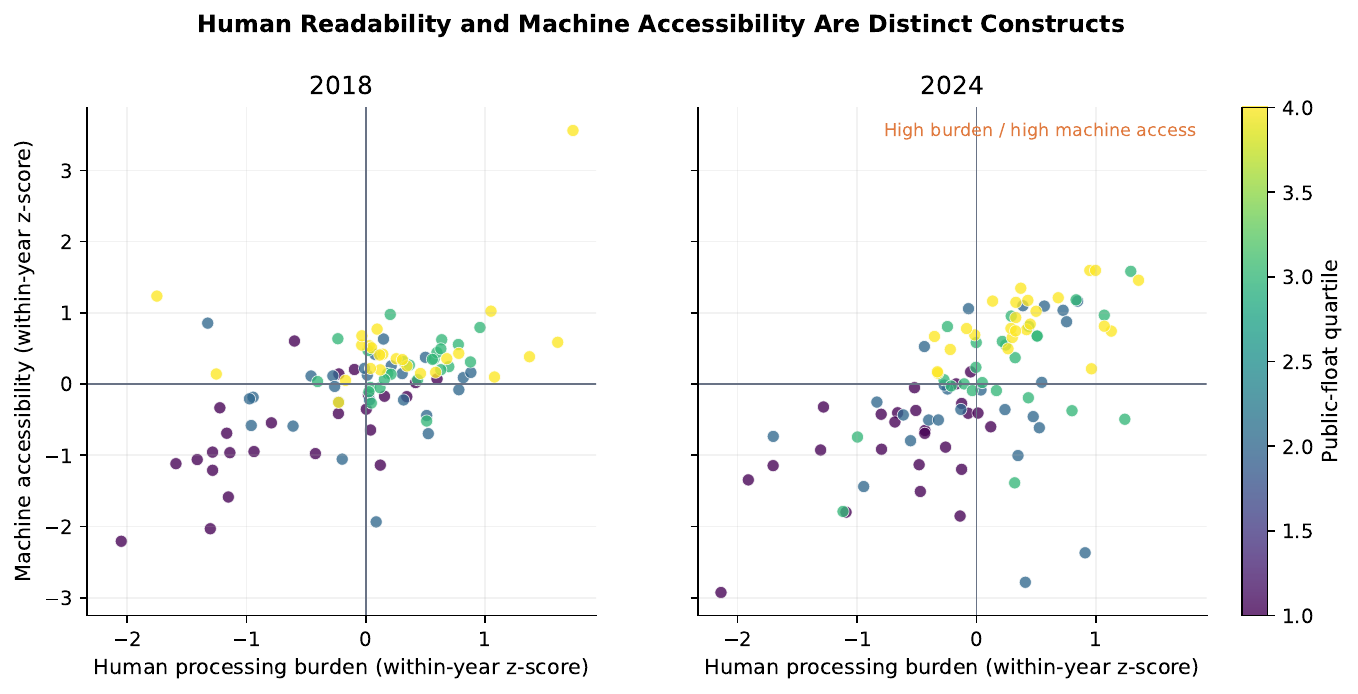}
\caption{Human processing burden and machine accessibility in a fixed balanced sample. Each point is a 10-K; colors denote within-year public-float quartiles. Both axes are equal-weight indices of transparent, within-year standardized primitives. The plotted relationship is descriptive.}
\label{fig:frontier}
\end{figure}

Figure \ref{fig:frontier} plots the two declared indices in 2018 and 2024. If conventional readability were a sufficient proxy for effective access, filings would lie near a downward-sloping locus: high human burden would imply low accessibility. They do not. Reports occupy all four quadrants, and public-float quartiles separate mainly along a positive diagonal. Larger issuers tend to produce longer and linguistically denser reports while supplying more standardized concepts and Inline XBRL tags.

The 2024 Spearman correlation between log document length and the machine-accessibility index is \ModernWordMachineRho{}. Positive correlation is not evidence that prose length makes algorithms smarter. Both attributes scale with issuer scope and reporting complexity. The scientific result is more basic: the same firm characteristic can increase human processing burden and machine-readable coverage. A scalar label such as ``readability'' loses information about which reader and which interface it describes.

Rank cross-classification reinforces the point. In each sample year, 35--40 percent of reports fall simultaneously above the median on human burden and machine accessibility, and another 20--30 percent occupy discordant quadrants. Hence neither index is a monotone relabeling of the other. This conclusion is robust to using primitive word count against facts, concepts, or tags rather than the composites.

\begin{figure}[H]
\centering
\includegraphics[width=.90\textwidth]{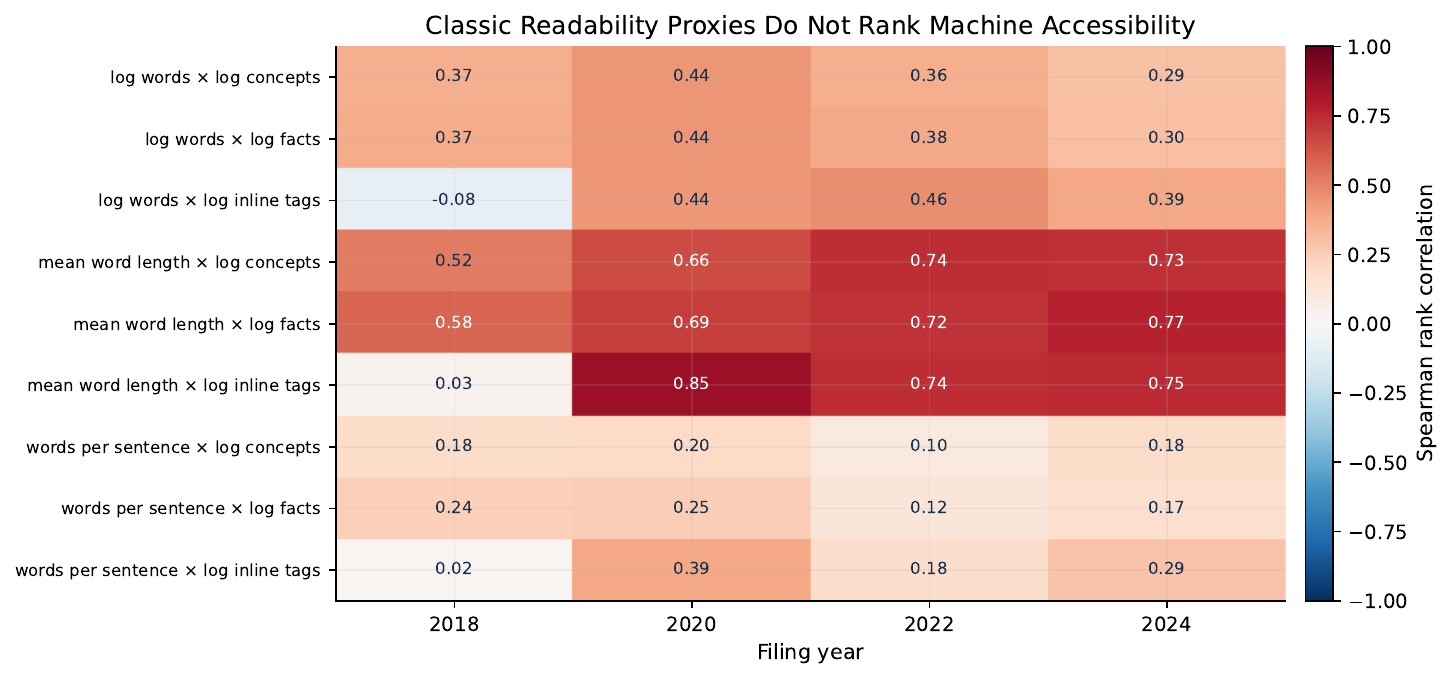}
\caption{Rank relationships between classic human-facing metrics and machine-access primitives. Cells report within-year Spearman correlations. Positive values mean that reports conventionally classified as more difficult can also be richer in machine-readable content. No correlation is interpreted causally.}
\label{fig:rankvalidity}
\end{figure}

\subsection{Criterion validation against realized machine performance}

An index assembled from convenient filing fields is not validated merely because its components sound machine related. We therefore replace the original equal-weight interpretation with an explicit criterion test. For each 2024 filing, we parse Inline XBRL and measure the standardized-tag share, concept uniqueness, duplicate-context incidence, conflicting duplicates, unit consistency, and standardized facts per thousand document words. These primitives are declared before looking at model outcomes and retain their natural direction: duplication and ambiguity reduce accessibility; standardization, uniqueness, unit consistency, and fact density increase it.

The behavioral criterion is filing-level grounded accuracy averaged across six models and the six realistic interfaces that require an operational retrieval or parsing step: BM25 HTML, TF--IDF reranking, table-aware HTML, full XBRL, retrieved XBRL, and hybrid access. Oracle arms are excluded because perfect localization would mechanically erase the filing property the index is intended to measure. A ridge model predicts the criterion from the structural primitives and word count under five-fold issuer-grouped cross-validation, so observations from an issuer never appear in both training and test folds.

\begin{table}[H]
\centering
\caption{Criterion Validation of Machine Accessibility}
\label{tab:constructvalidation}
\small
\begin{tabular}{lr}
\toprule
Criterion & Estimate \\
\midrule
Filings & 97 \\
Issuer-grouped cross-validated $R^2$ & 0.193 [0.059, 0.284] \\
Issuer-grouped cross-validated MAE & 0.082 [0.071, 0.093] \\
Structural index--behavior Spearman $\rho$ & 0.178 [-0.025, 0.379] \\
Word count--behavior Spearman $\rho$ & -0.047 \\
\midrule
HTML-interface held-out $R^2$ & 0.207 \\
XBRL-interface held-out $R^2$ & 0.011 \\
Hybrid-interface held-out $R^2$ & -0.022 \\
\bottomrule
\end{tabular}
\begin{minipage}{.94\textwidth}\footnotesize Notes: This is an initial criterion validation in the standardized numerical filing-use domain. Brackets are 95\% filing-bootstrap intervals formed from held-out prediction pairs. Predictions are out of sample under five-fold issuer-grouped cross-validation. Leave-one-model-out results are reported in the replication files.\end{minipage}
\end{table}

The structural index has Spearman correlation \ConstructBehaviorRho{} with realized grounded accuracy, compared with \WordBehaviorRho{} for word count. The multivariate primitives achieve out-of-sample $R^2=\ConstructCVRtwo$ (filing-bootstrap 95\% interval 0.059--0.284) and mean absolute error \ConstructCVMAE{}. Interface-family prediction is asymmetric: held-out $R^2$ is \ConstructHTMLRtwo{} for HTML but only \ConstructXBRLRtwo{} for XBRL and \ConstructHybridRtwo{} for hybrid access. Leave-one-model-out rank correlations remain positive for every held-out implementation (0.35--0.49), although level $R^2$ is sometimes negative. We therefore describe this as \emph{initial criterion validation} for standardized numerical filing use, not proof of a universal latent trait. The exercise converts machine accessibility from a convenient descriptive composite into a falsifiable measure while revealing where its predictive validity does and does not travel.

\begin{figure}[H]
\centering
\includegraphics[width=.72\textwidth]{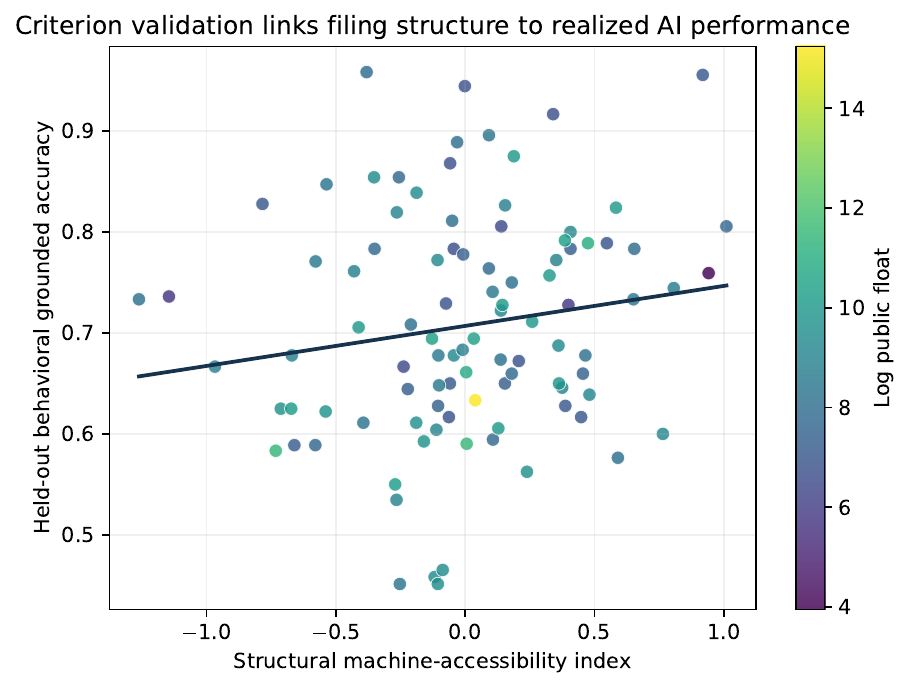}
\caption{Structural machine accessibility and realized multi-model performance. The behavioral criterion averages grounded accuracy over six realistic interfaces; color reports issuer public float. Predictive statistics use issuer-grouped held-out folds.}
\label{fig:constructvalidation}
\end{figure}

\subsection{Issuer scale and infrastructure fairness}

We estimate, for filing $j$ in year $t$ and two-digit industry $s$,
\begin{equation}
Y_{jts}=\alpha+\sum_{q=2}^{4}\beta_q\mathbf 1(Q_j=q)+\lambda_t+\gamma_s+\varepsilon_{jts},
\label{eq:modernscale}
\end{equation}
where $Q$ is the public-float quartile and $Y$ is a primitive or composite access measure. HC1 standard errors quantify sampling uncertainty within the fixed diagnostic sample; they do not repair nonrandom selection into public-company size.

Relative to the smallest quartile, the largest quartile has \ModernQFourMachineGap{} standard deviations higher machine accessibility (SE \ModernQFourMachineGapSE{}), conditional on year and industry. The corresponding human-burden difference is 1.12 standard deviations (SE 0.09). In primitives, the largest quartile has 0.83 log points more standardized fact observations and 2.25 log points more Inline tags. These gradients are precisely estimated within the sample and visible in cell medians.

\begin{table}[H]
\centering
\caption{Issuer Scale, Human Burden, and Machine Accessibility}
\label{tab:modernscale}
\begin{threeparttable}
\begin{tabular}{lrrr}
\toprule
Outcome & Q2 vs Q1 & Q3 vs Q1 & Q4 vs Q1 \\
\midrule
Machine accessibility index & 0.408 (0.117) & 0.959 (0.108) & 1.458 (0.108) \\
Human-burden index & 0.554 (0.100) & 0.945 (0.089) & 1.123 (0.095) \\
Interface-divergence index & 0.962 (0.187) & 1.904 (0.170) & 2.581 (0.182) \\
Log fact observations & 0.280 (0.071) & 0.622 (0.063) & 0.829 (0.057) \\
Log Inline tags & 0.788 (0.287) & 1.482 (0.279) & 2.251 (0.274) \\
\midrule
Year fixed effects & Yes & Yes & Yes \\
Two-digit SIC fixed effects & Yes & Yes & Yes \\
Observations & 400 & 400 & 400 \\
\bottomrule
\end{tabular}
\begin{tablenotes}[flushleft]\footnotesize
\item Notes: Each row is one OLS regression using the balanced 10-K sample. Coefficients compare public-float quartiles with Q1. HC1 standard errors are in parentheses. Composite indices use within-year standardized primitives. The regressions describe conditional scale gradients and are not causal effects of firm size.
\end{tablenotes}
\end{threeparttable}
\end{table}

\begin{figure}[H]
\centering
\includegraphics[width=.98\textwidth]{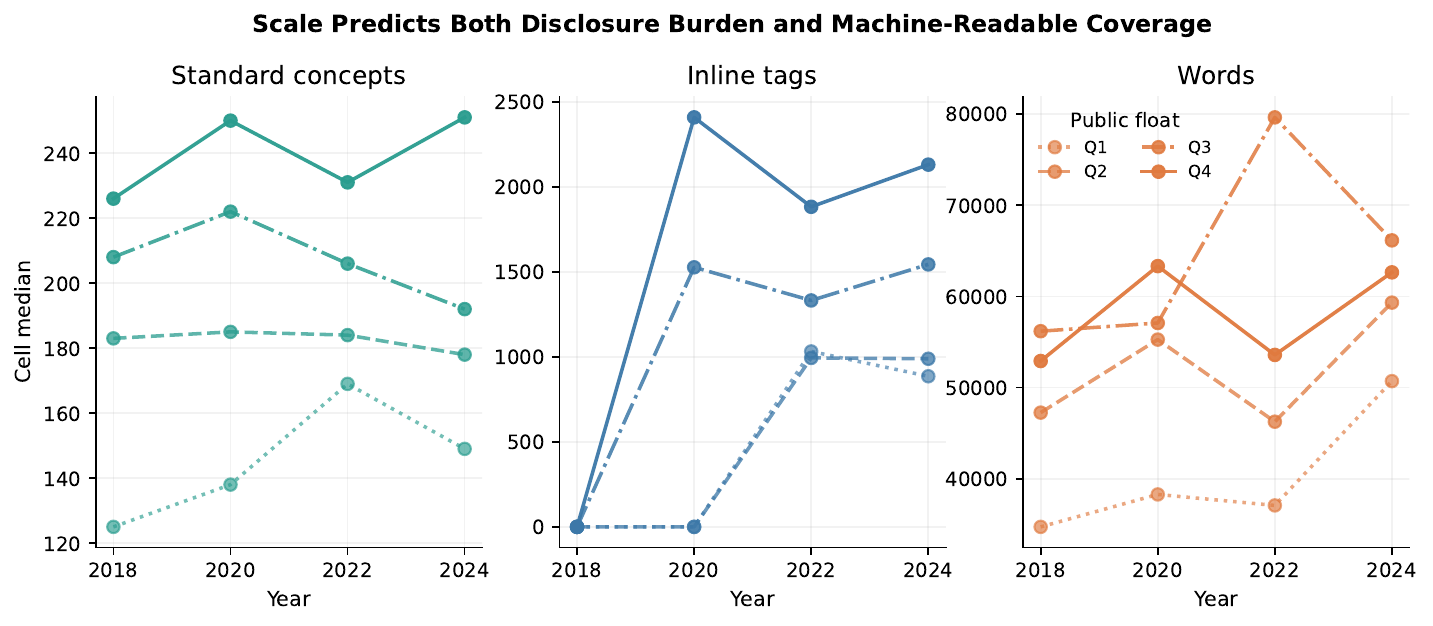}
\caption{Primitive disclosure measures by filing year and public-float quartile. Points are cell medians in the balanced 10-K sample. Standardized concepts and Inline tags proxy machine-ready coverage; words proxy human information quantity, not comprehension.}
\label{fig:sizegradients}
\end{figure}

The result qualifies two opposing narratives. Formal standardization does not eliminate scale-related differences in machine-readable coverage. Yet the gradient does not prove that small-issuer investors are disadvantaged overall: their reports are shorter, third-party tools may equalize access, and the marginal value of another tagged fact can differ by issuer complexity. We therefore call the evidence an infrastructure disparity, not a welfare inequality.

\subsection{The interface-divergence wedge}

\begin{figure}[H]
\centering
\includegraphics[width=.78\textwidth]{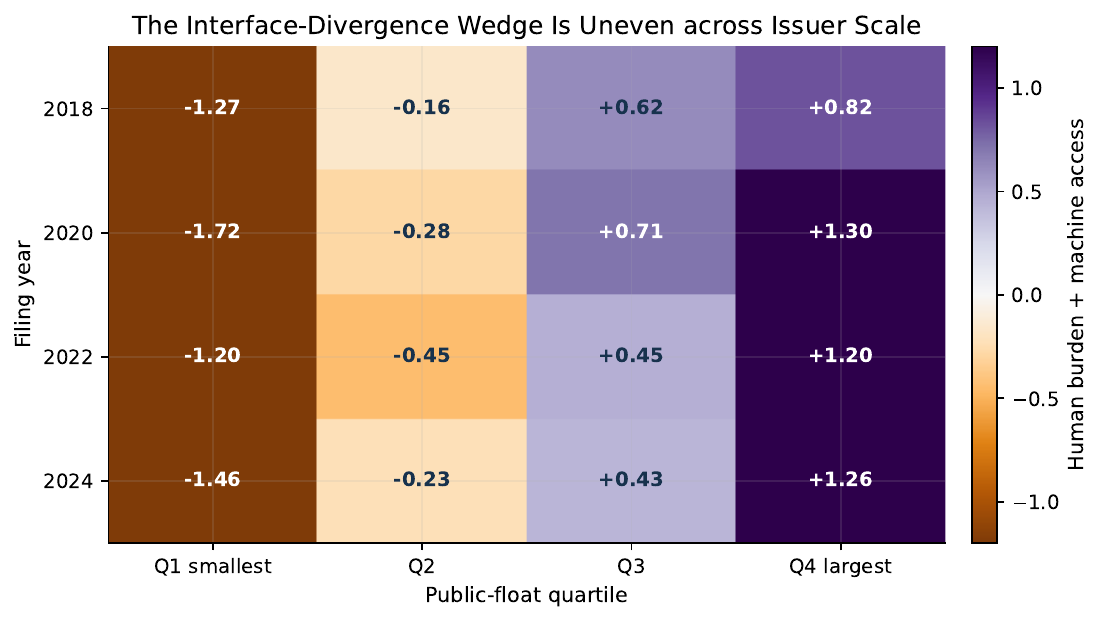}
\caption{Interface divergence by year and issuer scale. Cells report the mean of human burden plus machine accessibility, both standardized within year. Positive cells combine relatively high direct-reading burden with relatively high machine-ready coverage.}
\label{fig:interfaceheat}
\end{figure}

Figure \ref{fig:interfaceheat} summarizes the joint pattern. The smallest quartile has negative average divergence in every sampled year; the largest quartile has positive divergence. By 2024, the filing architecture is uniformly Inline XBRL, yet the cross-sectional wedge persists. Universal format adoption therefore does not make machine accessibility uniform, and denser structured coverage does not make the document easier for a person to read.

This pattern supplies the modern counterpart to the historical traffic decomposition. Historically, the measured attention effect changes when requests are separated by interface or automated-user rule. In the modern data, the measured accessibility of the same report changes when the intended reader shifts from unaided human to machine. Together the modules reject the maintained assumption that a stable one-dimensional proxy spans both regimes. They do not identify a single causal effect of ``AI'' because AI adoption, downstream audiences, and model outputs are not observed.

\subsection{A validity matrix for empirical research}

The evidence suggests a practical measurement matrix. Direct primary-document IPs are most defensible for source-level human-oriented acquisition, subject to shared-IP and deduplication error. XML/API requests are most defensible for source-level machine acquisition, but poor for downstream audience size. Word and syntax measures characterize human-facing burden, but do not measure structured extraction cost. Fact, concept, and tag coverage characterize machine-ready infrastructure, but do not measure model accuracy or investor comprehension. Market outcomes can reveal aggregate consequences but cannot by themselves assign the processing channel.

No row in this matrix is a universal ``investor attention'' measure. Researchers should name the stage they observe---source acquisition, transformation, distribution, cognition, or decision---and avoid carrying a coefficient interpretation across stages without validation. This recommendation becomes more important, not less, when AI systems hide intermediate processing behind a conversational interface.

\section{A Paired Experiment on Machine Processing Load}
\label{sec:factorial}

\subsection{Mechanism question and frozen design}

A preliminary four-arm comparison showed that an operational XBRL bundle could outperform conventional HTML retrieval, but the bundled design could not identify which component of machine processing load changed: XBRL simultaneously altered retrieval success, table parsing, representation, context length, and irrelevant information. The present paired factorial design separates these components. Every question--filing pair is evaluated under every interface. Pairing removes question composition as an explanation, while deliberately constructed interfaces isolate evidence localization, representation semantics, and context noise. The experiment does not measure human cognitive effort; it tests the parallel theoretical claim that processing difficulty is local to a reader, task, and interface.

The frozen pool contains 432 accession-grounded questions from 100 sampled reports, corresponding to 97 unique accessions, and five task families: assets, net income, net margin, current ratio, and liabilities to assets. Each of six hosted model implementations receives all 5,184 question--interface cells at temperature zero, for \FactorialCells{} model responses. The models are Gemini 2.5 Flash, Gemini 2.5 Pro, DeepSeek V3.2, Qwen 3.7 Plus, Qwen3-235B-A22B, and Doubao Seed 1.8. Canonical and returned model identifiers, raw responses, token accounting, latency, retry count, and dispatch order are retained in the non-public audit record; the manuscript reports provider-neutral model names. The task file and scoring contract were frozen before the formal run; the experiment was not publicly preregistered. Two implementations from each of the Gemini and Qwen model families are retained deliberately as within-family checks, so six implementations should not be misdescribed as six statistically independent model families.

The twelve interfaces form four mechanism families. First, a no-context control tests appropriate abstention. Second, five human-document interfaces vary localization quality: BM25 retrieval, TF--IDF reranking, table-aware retrieval, oracle text containing exactly the answer-bearing facts, and oracle text padded with real same-filing distractors. Third, four structured interfaces separate representation from retrieval: full XBRL, retrieved XBRL, oracle XBRL, and oracle XBRL padded to the text-noise arm's word budget. A compact flat-JSON arm removes XBRL taxonomy semantics while preserving field structure. Finally, a hybrid arm combines table-aware text and structured facts. Oracle text and oracle XBRL contain the same economic quantities; the two padded oracle arms contain the same target facts and nearly identical context budgets. Thus their paired differences identify representation effects under far stronger controls than an XBRL--HTML bundle comparison.

\begin{table}[H]
\centering
\caption{Factorial Interface Map and Ex Ante Evidence Coverage}
\label{tab:interfacemap}
\scriptsize
\setlength{\tabcolsep}{3.5pt}
\begin{tabular}{p{.21\textwidth}p{.27\textwidth}rrp{.22\textwidth}}
\toprule
Interface & Mechanism varied & Mean words & Coverage & Interpretation \\
\midrule
No context & Prior knowledge / abstention & 0 & 0.0\% & Leakage control \\
BM25 HTML & Lexical retrieval & 1,686 & 82.4\% & Baseline text bundle \\
TF--IDF + rerank & Alternative lexical ranking & 1,686 & 83.3\% & Retriever robustness \\
Table-aware HTML & Row and cell preservation & 1,363 & 83.6\% & Table reconstruction \\
Oracle text & Perfect fact localization & 14 & 100.0\% & Text upper bound \\
Oracle text + noise & Localization plus real distractors & 1,678 & 100.0\% & Text context-noise test \\
Flat JSON & Generic key--value structure & 5 & 100.0\% & Structure without taxonomy \\
Full XBRL & Complete structured filing facts & 5,373 & 100.0\% & Realistic structured bundle \\
Retrieved XBRL & Structured lexical retrieval & 133 & 97.7\% & Structured retrieval loss \\
Oracle XBRL & Perfect structured localization & 14 & 100.0\% & Fact-matched semantic test \\
Oracle XBRL + noise & Word-matched real distractors & 1,678 & 100.0\% & Structured noise test \\
Hybrid & Table-aware HTML plus facts & 1,516 & 99.8\% & Complementary interfaces \\
\bottomrule
\end{tabular}
\begin{minipage}{.96\textwidth}\footnotesize Notes: Coverage is the share of 432 questions for which the supplied blocks jointly contain every ground-truth input fact under the strict co-occurrence rule. Mean words use whitespace tokenization for transparent interface matching; API tokenizer counts differ.\end{minipage}
\end{table}

The design yields five falsifiable predictions. \emph{H1:} better evidence localization improves grounded accuracy within HTML. \emph{H2:} after holding target facts and context fixed, the residual XBRL-semantic advantage is substantially smaller than the raw XBRL--retrieved-HTML gap. \emph{H3:} if taxonomy semantics add information beyond generic structure, oracle XBRL outperforms flat JSON. \emph{H4:} irrelevant context reduces accuracy within both text and structured representations. \emph{H5:} interface effects replicate in sign across hosted implementations rather than depending on one deployment.

\subsection{Strict evidence scoring and estimands}

Every response must be a JSON object containing a value, unit, abstention decision, and an array of evidence identifiers. Grounded accuracy requires valid JSON, the correct unit, a value within the ex ante tolerance, and citations that jointly support every input fact needed for the answer. This last condition is stricter than checking whether a cited identifier merely appeared in the prompt. For a ratio, for example, at least one cited block must support the numerator and at least one must support the denominator. If retrieval omits a needed fact, a memorized or lucky numerical answer cannot pass.

For interfaces $a$ and $b$, the primary estimand is the within-question, within-model difference
\begin{equation}
 \widehat\Delta_{ab}=\frac{1}{Q}\sum_{q=1}^{Q}
 \left(\frac{1}{M}\sum_{m=1}^{M}[Y_{qma}-Y_{qmb}]\right),
\end{equation}
where $Y$ is grounded correctness, $Q=432$, and $M=6$. Pairing remains at the question--model level, but inference accounts for dependence among questions from the same filing. We average implementation-specific differences within question, compute accession-clustered standard errors over 97 unique accessions, and report filing-level bootstrap confidence intervals that resample accessions with all their questions. We also report every contrast separately by model, task family, and issuer-size quartile. This estimand is intentionally narrower than investor welfare: it measures how an assigned representation changes grounded machine processing of the same disclosed facts.

\subsection{Evidence localization drives the interface wedge in standardized numerical tasks}

\begin{table}[H]
\centering
\caption{Paired Multi-Model Interface Experiment}
\label{tab:factorial}
\small
\begin{tabular}{lrrr}
\toprule
Interface & Accuracy & Model SD & Mean context words \\
\midrule
No context & 0.0\% & 0.0 & 0 \\
HTML: BM25 & 47.2\% & 5.8 & 1,686 \\
HTML: TF-IDF + rerank & 49.9\% & 4.8 & 1,686 \\
HTML: table-aware & 58.8\% & 5.0 & 1,363 \\
Oracle text & 99.5\% & 0.5 & 14 \\
Oracle text + noise & 99.8\% & 0.2 & 1,678 \\
Flat JSON & 99.3\% & 0.7 & 5 \\
Full XBRL & 94.2\% & 4.1 & 5,373 \\
Retrieved XBRL & 82.4\% & 12.5 & 133 \\
Oracle XBRL & 99.6\% & 0.5 & 14 \\
Oracle XBRL + noise & 99.8\% & 0.6 & 1,678 \\
Hybrid & 91.7\% & 2.6 & 1,516 \\
\bottomrule
\end{tabular}
\begin{minipage}{.94\textwidth}\footnotesize Notes: Each of 432 question--filing pairs from 100 sampled reports, corresponding to 97 unique accessions, is evaluated under every interface by six hosted implementations. Accuracy is averaged across implementations; model SD reports cross-implementation dispersion. Paired contrast inference is accession clustered and reported separately.\end{minipage}
\end{table}

For the standardized numerical filing tasks studied here, the paired mechanism decomposition substantially narrows the interpretation of the initial bundled comparison. Grounded accuracy rises from \FactorialBM\% under BM25 HTML to \FactorialOracleText\% when the same text facts are perfectly localized. Oracle XBRL reaches \FactorialOracleXBRL\%, but its paired advantage over oracle text is only \FactorialSemanticEffect{} percentage points. Replacing the XBRL representation with flat JSON changes accuracy by only \FactorialTaxonomyEffect{} points. The economically large wedge is therefore not an intrinsic inability of language models to read prose, nor a large standalone return to taxonomy labels. In this task environment, it is principally a localization and table-reconstruction problem created upstream of generation.

The hypothesis accounting is deliberately two sided. H1 is supported: oracle localization exceeds BM25 by 52.3 percentage points (accession-clustered SE 2.25 points; 95\% CI [47.82, 56.74]), and table-aware HTML closes 11.6 points of the raw gap. H2 is supported: the fact-matched XBRL--text contrast is 0.08 points (clustered 95\% CI [\FactorialSemanticCILow, \FactorialSemanticCIHigh]). H3's directional prediction is not supported: XBRL exceeds flat JSON by 0.23 points (clustered 95\% CI [\FactorialTaxonomyCILow, \FactorialTaxonomyCIHigh]). H4 is also not supported in the declared direction: oracle-minus-padded effects are $-0.31$ points for text (95\% CI [$-0.65$, 0.03]) and $-0.19$ for XBRL (95\% CI [$-0.51$, 0.12]). H5 is supported descriptively: every implementation exhibits a large localization gap, while no implementation exhibits a material oracle-XBRL advantage. We do not treat six hosted implementations as a random sample of all future models.

\begin{table}[H]
\centering
\caption{Accession-Clustered Representation Contrasts and Practical Equivalence}
\label{tab:equivalence}
\small
\begin{tabular}{lrrrr}
\toprule
Contrast & Effect (pp) & Clustered 95\% CI & Filing-bootstrap 95\% CI & TOST $p$ \\
\midrule
Oracle XBRL $-$ oracle text & 0.08 & [-0.25, 0.40] & [-0.23, 0.39] & <0.001 \\
Oracle XBRL $-$ flat JSON & 0.23 & [-0.16, 0.62] & [-0.15, 0.61] & <0.001 \\
\bottomrule
\end{tabular}
\begin{minipage}{.96\textwidth}\footnotesize Notes: Effects preserve question-level pairing and average the six implementation-specific differences within question. Standard errors cluster by 97 unique accessions; bootstrap intervals resample accessions with all their questions (10,000 draws). TOST is a post hoc robustness analysis using a symmetric $\pm1$ percentage-point practical-equivalence margin. Both 90\% confidence intervals lie within that margin.\end{minipage}
\end{table}

The practical-equivalence analysis is explicitly post hoc. With a symmetric $\pm1$ percentage-point margin, both accession-clustered 90\% confidence intervals fall inside the equivalence region, and both TOST tests reject non-equivalence. The 95\% confidence intervals and filing-bootstrap intervals lead to the same substantive conclusion. The matched-evidence design therefore rules out a practically material representation advantage under this declared robustness threshold; this is stronger than interpreting a nonsignificant difference as equality.

\begin{figure}[H]
\centering
\includegraphics[width=.96\textwidth]{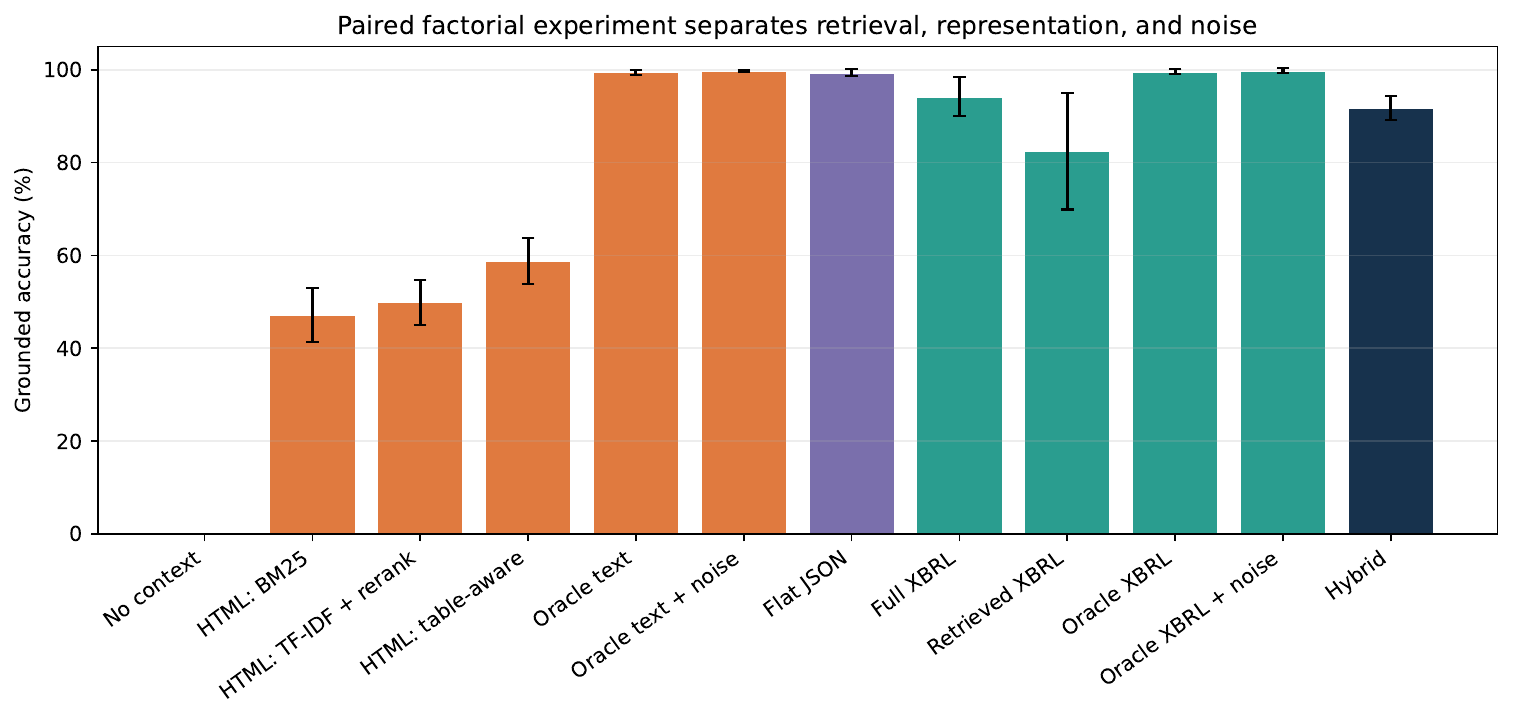}
\caption{Grounded accuracy across twelve paired interfaces. Bars average model-specific accuracy; whiskers show cross-model standard deviations. Every interface contains the same 432 questions.}
\label{fig:factorialaccuracy}
\end{figure}

\begin{figure}[H]
\centering
\includegraphics[width=.96\textwidth]{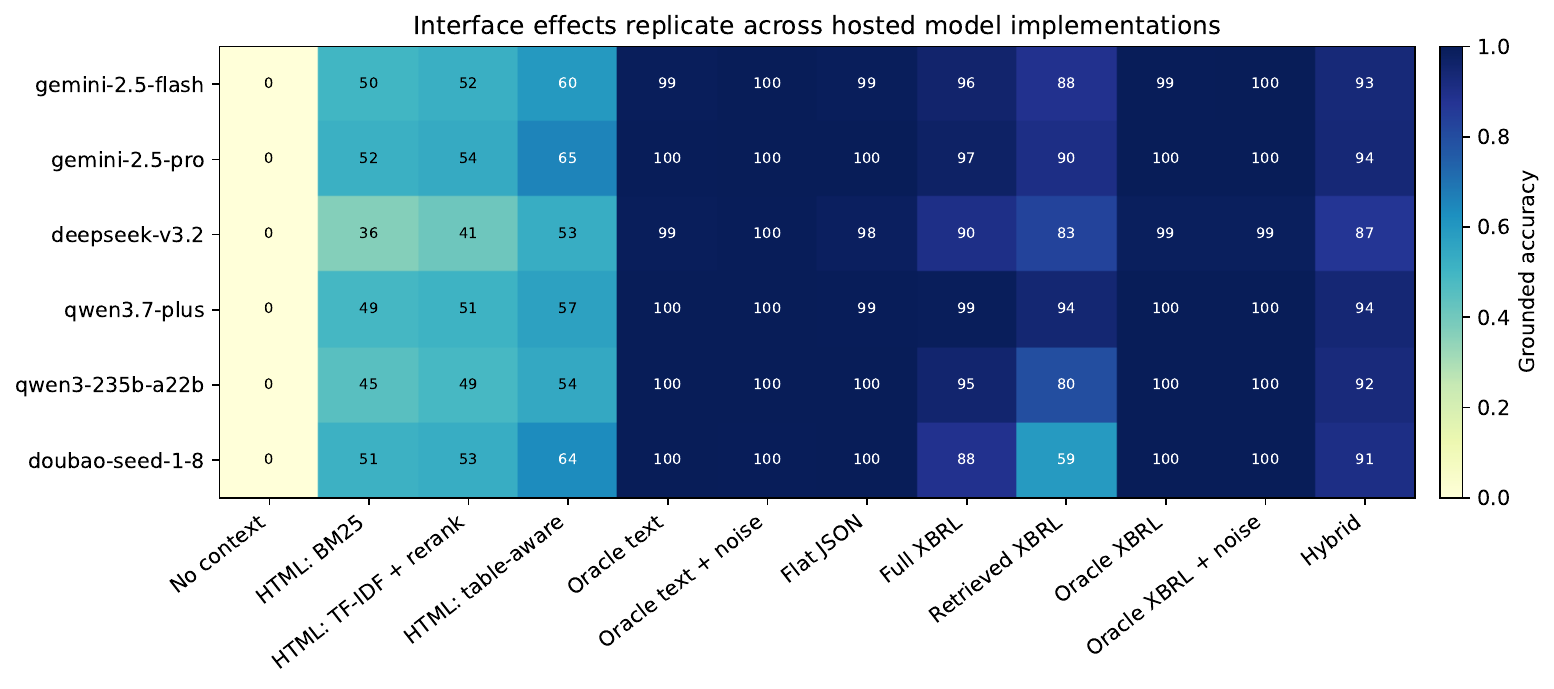}
\caption{Grounded accuracy by hosted model implementation and interface. The complete heat map prevents a pooled mean from concealing model dependence.}
\label{fig:factorialheatmap}
\end{figure}

\begin{figure}[H]
\centering
\includegraphics[width=.88\textwidth]{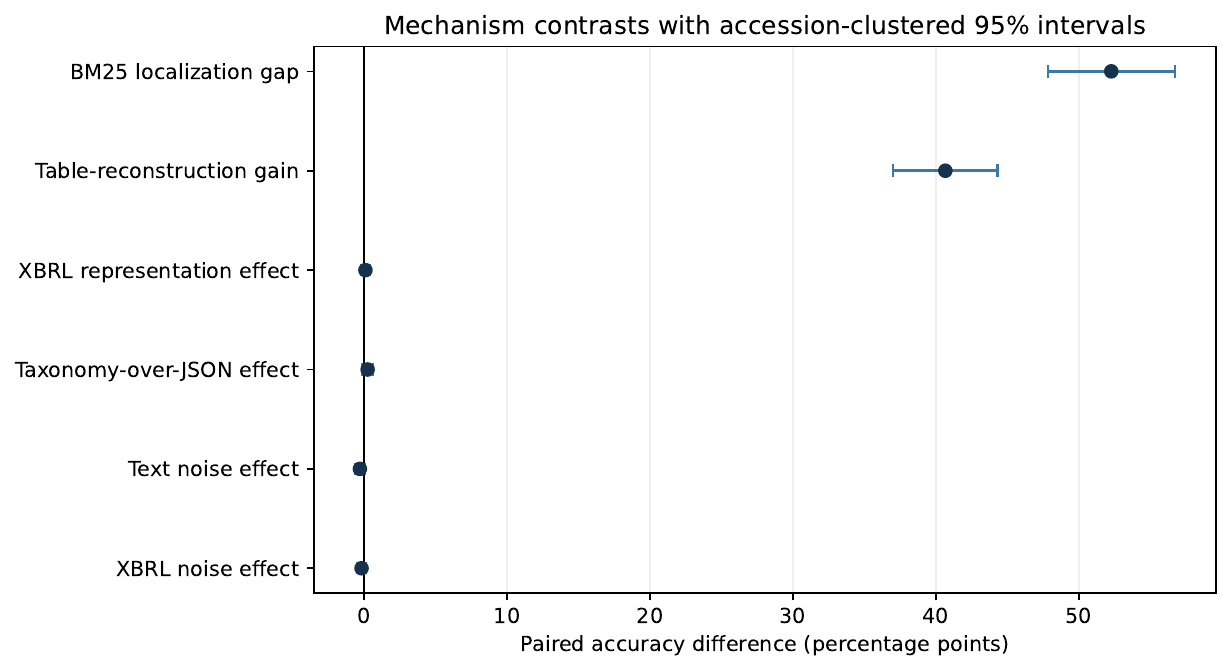}
\caption{Paired mechanism contrasts with accession-clustered 95\% confidence intervals. Question-level pairing is preserved while dependence within filing is reflected in inference.}
\label{fig:factorialcontrasts}
\end{figure}

\subsection{Operational failure decomposition}

\begin{table}[H]
\centering
\caption{Failure Decomposition across Operational Interfaces}
\label{tab:failuredecomposition}
\footnotesize
\begin{minipage}{.96\textwidth}
\centering
\textit{Panel A1. Evidence and value failures}\par\smallskip
\begin{tabular*}{\textwidth}{@{\extracolsep{\fill}}lrrrr@{}}
\toprule
Interface & Evidence absent & Unsupported value & Wrong unit & Value/calculation \\
\midrule
BM25 HTML & 17.6\% & 2.2\% & 0.0\% & 8.0\% \\
Table-aware HTML & 16.4\% & 1.9\% & 0.0\% & 17.5\% \\
Full XBRL & 0.0\% & 1.5\% & 0.2\% & 2.3\% \\
Retrieved XBRL & 2.3\% & 0.9\% & 0.0\% & 2.2\% \\
Hybrid & 0.2\% & 0.7\% & 0.0\% & 3.2\% \\
\bottomrule
\end{tabular*}
\par\medskip
\textit{Panel A2. Citation and response-control outcomes}\par\smallskip
\begin{tabular*}{\textwidth}{@{\extracolsep{\fill}}lrrrr@{}}
\toprule
Interface & Incomplete citation & Abstention & Schema error & Grounded correct \\
\midrule
BM25 HTML & 2.4\% & 22.5\% & 0.0\% & 47.2\% \\
Table-aware HTML & 1.9\% & 3.4\% & 0.1\% & 58.8\% \\
Full XBRL & 1.1\% & 0.6\% & 0.1\% & 94.2\% \\
Retrieved XBRL & 0.8\% & 10.7\% & 0.6\% & 82.4\% \\
Hybrid & 1.5\% & 2.6\% & 0.0\% & 91.7\% \\
\bottomrule
\end{tabular*}
\par\medskip
\textit{Panel B. Coverage, numerical accuracy, and support validity}\par\smallskip
\begin{tabular*}{\textwidth}{@{\extracolsep{\fill}}lrrrr@{}}
\toprule
Interface & Evidence coverage & Numerical accuracy & Grounded $\mid$ coverage & Correct value, invalid support \\
\midrule
BM25 HTML & 82.4\% & 57.6\% & 57.3\% & 10.4\% \\
Table-aware HTML & 83.6\% & 69.5\% & 70.4\% & 10.6\% \\
Full XBRL & 100.0\% & 95.3\% & 94.2\% & 1.1\% \\
Retrieved XBRL & 97.7\% & 84.0\% & 84.4\% & 1.5\% \\
Hybrid & 99.8\% & 93.4\% & 92.0\% & 1.7\% \\
\bottomrule
\end{tabular*}
\end{minipage}
\par\smallskip
\begin{minipage}{.97\textwidth}\footnotesize Notes: Rates average 432 questions across six hosted implementations. Categories in the upper panel are mutually exclusive and exhaustive apart from an unreported residual that is zero. Evidence absent means at least one required input fact is missing ex ante. Unsupported value means evidence is available, the returned value is wrong, and citations do not cover all required facts. Categories identify observable control failures, not latent reasoning processes. Numerical accuracy ignores citation validity but requires valid JSON, a non-abstaining response, the correct unit, and a value within tolerance.\end{minipage}
\end{table}

Table \ref{tab:failuredecomposition} makes the distinction in Equation \ref{eq:three_layers} observable. Full XBRL contains every required input fact, yet grounded accuracy is 94.2\%, because complete availability does not eliminate unsupported value selection, calculation, abstention, or citation failures. Hybrid access has 99.8\% coverage but 91.7\% grounded accuracy. Retrieved XBRL's largest residual failure is abstention (10.7\%), not missing evidence alone (2.3\%). Table-aware HTML changes the failure mix rather than merely adding evidence: relative to BM25, abstention falls from 22.5\% to 3.4\%, while more attempted answers expose value or calculation errors. These mutually exclusive categories identify observable control failures, not latent reasoning processes; they locate failures at retrieval, evidence use, citation, and output-contract stages.

This finding is more scientifically restrictive, and therefore more useful, than a generic claim that ``XBRL helps AI.'' For standardized numerical filing tasks, standardization has high value when it makes relevant accounting facts addressable across accession, reporting period, unit, and segment and when it makes tables reconstructable. Once those services are supplied equally, the tested representation contrasts are practically equivalent under the post hoc threshold. AIS evaluation should consequently measure evidence coverage, accounting-context resolution, grounded use, and provenance separately instead of treating file-format compliance as a sufficient reliability control.

\subsection{Counterfactual evidence, decoys, and memorization}

Public-company facts may have entered model training data, so correct answers on historical filings do not by themselves prove evidence use. We draw a task-stratified 100-question subset and deterministically replace every target value. For calculations, answers are recomputed from the perturbed inputs. The original values then reappear only as plausible prior-period, wrong-unit, or segment-level decoys. Each model receives no context, oracle counterfactual text, oracle counterfactual XBRL, and the corresponding two decoy conditions. One deterministic draw implied a liabilities-to-assets ratio above 6,000 percent and repeatedly produced provider nonresponse; before final analysis we exclude that question uniformly across all five interfaces and all six models rather than selectively delete failed cells. The complete paired analysis therefore contains \CounterfactualCells{} responses from 99 questions.

\begin{table}[H]
\centering
\caption{Counterfactual Evidence and Decoy Test}
\label{tab:counterfactual}
\small
\begin{tabular}{lrr}
\toprule
Interface & Grounded accuracy & Model SD \\
\midrule
No context (Appropriate abstention) & 100.0\% & 0.0 \\
Counterfactual text (Grounded accuracy) & 99.7\% & 0.5 \\
Counterfactual XBRL (Grounded accuracy) & 99.7\% & 0.8 \\
Text + prior-value decoys (Grounded accuracy) & 99.5\% & 0.6 \\
XBRL + prior-value decoys (Grounded accuracy) & 99.5\% & 0.6 \\
\bottomrule
\end{tabular}
\begin{minipage}{.94\textwidth}\footnotesize Notes: One hundred accession-grounded questions were deterministically perturbed. One question was excluded uniformly across models and interfaces because the deterministic draw implied a liabilities-to-assets ratio above 6,000 percent and repeatedly triggered provider nonresponse. Original values reappear only as plausible prior-period, unit, or segment decoys.\end{minipage}
\end{table}

The test does not fabricate a fully internally consistent financial statement and is not described as one. It is a counterfactual evidence packet designed for a narrower diagnostic: whether the model follows supplied answer-bearing evidence when a familiar issuer value is available as a tempting alternative. Counterfactual-text and counterfactual-XBRL accuracy are \CounterfactualText\% and \CounterfactualXBRL\%; after adding the original-value decoys, they are \CounterfactualTextDecoy\% and \CounterfactualXBRLDecoy\%. The comparison directly bounds the extent to which the main interface findings can be explained by memorized filing values.

\begin{figure}[H]
\centering
\includegraphics[width=.82\textwidth]{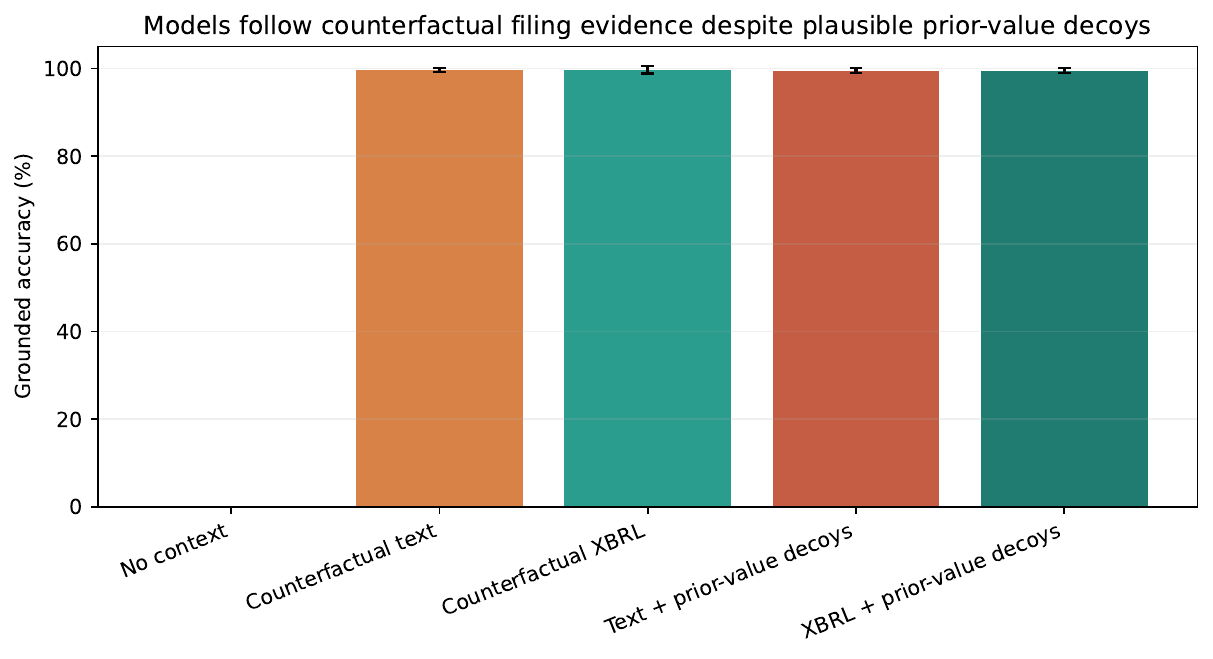}
\caption{Grounded use of deterministically perturbed filing evidence, with original values retained as plausible decoys.}
\label{fig:counterfactual}
\end{figure}

\subsection{Boundary of the machine experiment}

These experiments provide causal evidence on interface mechanisms for grounded model outputs under one frozen response contract, not effects on investor beliefs, trades, or welfare. They cover standardized numerical extraction and calculation, where exact ground truth and required evidence can be declared ex ante. Narrative synthesis, accounting-policy interpretation, and human reliance are outside the identified domain. This boundary is central to the paper's measurement argument: evidence about one reader, task, and interface should not be silently relabeled as evidence about another.

\section{Extensions and Robustness}

\subsection{Robot definitions are economic definitions}

The preferred 50-request cutoff approximates lower-volume acquisition, but the 1,000-request and SEC-flag-only outcomes retain progressively more automation. Comparing them asks whether the result belongs to direct visitors or to high-volume information infrastructure. A stable coefficient would indicate insensitivity to this distinction; divergence is itself evidence that the mandate changed automated retrieval.

We do not choose the filter that produces the most favorable coefficient. All three definitions are generated in one pass over the logs and reported as an outcome family with FDR adjustment. This approach is particularly important for XBRL because automated users are not contamination in the ordinary sense: they are among the intended beneficiaries.

\subsection{Predetermined complexity}

Appendix Figure \ref{fig:heterogeneity} reports high-minus-low interactions for document length, sentence length, numeric density, and table intensity. They do not uniformly support generic ``cognitive relief.'' The nonuniform pattern is instead compatible with an interface-specific mechanism in which tags structure numerical statements more directly than narrative prose.

\subsection{Placebos and continuity}

We assign two false policy dates, December 15, 2008 and March 15, 2009, and use only observations before the real rule date. Neither placebo produces a statistically distinguishable total-attention effect. The tests are not powerful proof of parallel trends, but they reduce concern that any arbitrary split generates the result.

Covariate-continuity regressions compare pre-policy document length, sentence length, numeric density, table count, and prior attention around the threshold. Table count exhibits a discontinuity of -53.7 tables ($p=0.003$); the other four tests do not reject at conventional levels. This imbalance is a direct warning against a strong local-randomization interpretation. Adding baseline-complexity trends leaves the total estimate near -0.111, whereas allowing two-digit-industry-specific post trends attenuates it to -0.075 (SE 0.118). Historical float recovery, industry composition, and sparse below-threshold support remain material limitations.

\subsection{Alternative bandwidths and multiplicity}

The \$1 billion bandwidth is locally appealing but has few firms and a weak first stage. The preferred \$2.5 billion bandwidth trades locality for usable support. The full candidate range improves precision but relies on extrapolation farther above the threshold. We show all three rather than selecting after observing the outcome.

Primary, channel, robot, dynamic, and heterogeneity estimates create multiple opportunities for chance findings. The machine-readable CSV reports raw $p$-values and Benjamini--Hochberg $q$-values within prespecified families. The paper's conclusion rests on coefficient patterns across interfaces, not one adjusted or unadjusted threshold.

\section{Implications for Cognitive-Load Research and Accounting Information Systems}

The empirical modules develop the cognitive-load question under a changed information environment. The claim is not that AI eliminates cognitive constraints. It is that technology relocates part of the constraint from unaided reading to acquisition, evidence transformation, verification, and integration. A disclosure can become easier on one margin and harder on another. This observation gives cognitive-load research a measurement rule: every claim about complexity should state the reader, task, interface, and processing stage it targets.

\subsection{Digital disclosure changes the information supply chain}

The usual disclosure model connects a firm directly to an investor. Structured data add intermediaries: filing agents, parsers, aggregators, terminals, screeners, and models. Once one request can serve many users, server traffic and effective distribution separate. This is not unique to XBRL; it applies to regulatory APIs, machine-readable contracts, and AI retrieval systems.

The empirical implication is that direct attention can be a misleading welfare statistic. A regulator observing fewer distinct IPs after introducing structured data might infer lower use, even though intermediaries are distributing the information more efficiently. Conversely, a surge in XML requests might reflect repetitive crawling rather than economically meaningful expansion. Interface composition and consolidation must be measured together.

\subsection{Implications for XBRL evaluation}

The evidence supports the claim that XBRL assets became part of the acquisition workflow. It provides less support for a universal reduction in investor cognitive load. The primary HTML document remains, narrative complexity remains, and a machine-readable request may never be seen by an end investor. Policy evaluation should therefore distinguish at least four outcomes: structured availability, direct structured retrieval, downstream distribution, and decision quality.

The modest first stage near the nominal threshold also matters. Eligibility rules interact with filer status, voluntary behavior, transition timing, and data measurement. Studies using the phase-in should verify filing-level treatment and report crossovers. A calendar dummy alone is not enough.

\subsection{Implications for attention and information-processing research}

EDGAR traffic is often called investor attention. Our results suggest more precise labels. Low-volume unique-IP retrieval is a proxy for direct information acquisition, not a direct measure of total cognitive effort. High-volume traffic is a mixture of automation, intermediation, and crawling. Asset-specific traffic reveals task choice. Researchers should align the label with the filtering rule and report whether IPs are deduplicated by day, accession, or event window.

This recommendation extends beyond EDGAR. Search volume, clicks, and platform views measure entry into a particular interface. When an AI assistant answers from cached, licensed, or retrieved information, the source may receive no contemporaneous user-level event. Time-series breaks in platform design can therefore be economically larger than ordinary measurement error. Validation should be repeated after a material change in distribution technology rather than assumed to transfer unchanged across technological regimes.

\subsection{Design and control principles for AI-enabled accounting information systems}

A regulator or platform serving both people and machines faces a multi-objective AIS design problem. Plain-language requirements target unaided comprehension; structured-data mandates target extraction and reuse; public APIs target distribution. Improvements on one interface need not spill over to another. Four control principles follow. First, systems should record evidence use rather than infer it from availability. Second, format compliance is not a sufficient reliability control: context resolution and table reconstruction must be tested. Third, every answer should satisfy an explicit evidence contract linking each input fact to a stable identifier. Fourth, provenance should be treated as an internal-control object---retaining accession, transformation, model, and evidence lineage so an answer can be reproduced and audited.

The scale gradient and experiment suggest a concrete monitoring agenda. Operators can publish filing-level diagnostics for standardized concept coverage, extension use, parsing errors, API latency, retrieval coverage, abstention, and grounded answer accuracy, stratified by issuer size. Making answer-bearing evidence directly addressable raises grounded accuracy by more than 50 percentage points relative to BM25 retrieval in this numerical task environment. Realistic structured interfaces achieve smaller gains because evidence selection, contextual ambiguity, and citation requirements remain operational constraints.

\subsection{Implications for information fairness}

AI intermediation can either narrow or widen information gaps. A low-cost public assistant may turn a 150-page report into accessible answers, relaxing literacy and time constraints. A proprietary system with superior data normalization, retrieval, and verification may instead amplify advantages enjoyed by sophisticated institutions. The relevant comparison is not ``human versus AI'' in the abstract but the joint distribution of raw-data coverage, model quality, verification cost, and access price.

Our results identify three reasons not to infer fairness from public availability alone. First, a source request understates the audience of an intermediary. Second, machine-ready coverage varies with issuer scale. Third, identical model access is not identical processing access: across six hosted implementations, BM25 HTML yields \FactorialBM\% grounded accuracy, while perfectly localized text yields \FactorialOracleText\%. Oracle XBRL adds only \FactorialSemanticEffect{} percentage points relative to fact-matched oracle text. The relevant disparity is therefore access to evidence-localization and table-reconstruction infrastructure, not merely possession of an XBRL file or a nominally public document. This is evidence of interface dependence, not discriminatory treatment or realized investor welfare.

\subsection{Implications for firms and intermediaries}

Firms may rationally optimize disclosure for multiple readers. More standardized facts can improve comparison and model retrieval, while clear narrative remains valuable for context, accountability, and verification. The positive association between human burden and machine accessibility is therefore not necessarily a failure of reporting. It may reflect the genuine scope of large firms. The actionable question is whether additional complexity conveys incremental information on each interface.

Intermediaries inherit a provenance obligation. When one filing request supports many downstream answers, mistakes can propagate at scale and become invisible in source logs. Systems that retain accession-level citations, distinguish reported from inferred values, and expose transformations make the observability wedge more auditable. Future disclosure research should measure these design choices directly.

\section{Limitations}

\noindent\textbf{Historical identification and traffic measurement.} Anonymized IPs are not people or institutions, source logs omit downstream redistribution, and crawler flags are incomplete. Public float is reconstructed retrospectively; the statutory first stage is modest, below-threshold support is sparse, and one baseline complexity measure is discontinuous. The historical module is therefore a measurement diagnostic, not a clean local-randomization or welfare design.

\noindent\textbf{Construct and sample scope.} Company Facts and the formal tasks emphasize standardized numerical information, not narrative synthesis or accounting-policy interpretation. The 400-document diagnostic sample is balanced rather than population weighted. Initial criterion validity is strongest for HTML-facing performance and does not establish a universal latent accessibility trait.

\noindent\textbf{System and implementation scope.} Six hosted implementations improve replication, but two pairs share model families, hosted systems can change over time, and all systems use one deterministic response contract and five task families. Oracle packets abstract from end-to-end production systems; counterfactual packets are targeted evidence tests, not complete synthetic financial statements. Formal tasks were frozen before execution but not publicly preregistered.

\noindent\textbf{Outcome scope.} Machine grounded accuracy is not human comprehension, trust, portfolio choice, price discovery, or welfare. The paper contains no survivorship-bias-free market merge and no randomized human-interface outcome. These are evidence boundaries, not effects inferred from the present data.

The paper also leaves equilibrium disclosure incentives and individual cognitive heterogeneity open. If managers learn how automated readers score language or select facts, they may optimize reports for algorithms rather than investors. If readers differ in financial literacy, working memory, or access to transformation tools, the same interface effect may translate into different realized burdens. Detecting either response requires implementation-change timing, participant-level outcomes, prompt-stable benchmarks, and pre-specified adversarial tests. We regard these as extensions, not claims established by the present evidence.

\section{Conclusion: Cognitive Load after the Interface Shift}

The processing environment of financial disclosure has changed. Reconstructing the first U.S. XBRL phase-in shows a modest statutory first stage, a decline or imprecise change in direct total acquisition depending on specification, weaker primary-document acquisition, and increased structured-file acquisition. The most defensible historical interpretation is suggestive interface reallocation combined with possible intermediary consolidation. This is the acquisition-stage extension of the cognitive-load question: a source request is not the same object as the information-processing activity it may initiate.

The modern evidence establishes that disclosure complexity is interface-local. Inline XBRL is now the modal architecture, public APIs expose standardized facts at scale, and large-firm reports combine greater human processing burden with substantially greater machine-ready coverage. Human readability and machine accessibility are not opposites on one axis. They are separate attributes whose joint distribution determines the processing environment faced by different readers and technologies.

The paired experiment identifies one transformation-stage mechanism. Across six hosted model implementations, ordinary HTML retrieval performs far below perfectly localized text. Accession-clustered confidence intervals and filing-level bootstrap intervals place the oracle-XBRL contrasts with fact-matched text and flat JSON inside a post hoc $\pm1$ percentage-point practical-equivalence region. Failure decomposition further shows why availability is insufficient: realistic interfaces can contain nearly every required fact yet still fail through unsupported value selection, calculation, abstention, or incomplete evidence citations. Counterfactual values and original-value decoys show that models follow supplied evidence rather than familiar issuer facts within this diagnostic. These are machine-processing results, not measures of human comprehension; their theoretical role is to reveal where an interface can add or remove transformation load.

The methodological conclusion is direct. Conditioning on realized adoption can make a regulatory design look cleaner than it is. Removing high-volume users can remove the mechanism under study. Calling every EDGAR request ``investor attention'' can conceal changes in the information supply chain. Calling a long filing ``inaccessible'' can conceal a rich structured layer. Calling a filing ``public'' can conceal differences in transformation capacity.

The paper does not show that traditional measures are useless. It shows that they are local to a reader, interface, task, and processing stage. In an AI-intermediated market, sound cognitive-load and disclosure research must test measurement invariance after technological breaks and treat interface design as part of the processing environment. Its three contributions are a reader--task--interface extension of cognitive-load theory, causal evidence on how evidence addressability affects grounded model outputs in standardized numerical tasks, and initial, task-conditioned criterion validation of transparent machine-accessibility measures. The result is an auditable distinction among information that is public, evidence that is addressable, and answers that are grounded.

\clearpage
\appendix
\section{Reconstruction Protocol}

\subsection{Raw sources and immutability}

The project stores the SEC bulk submissions archive, company facts archive, daily log ZIP files, and primary filing documents without altering their contents. Download metadata and checksums are retained where available. Intermediate tables are regenerated by deterministic scripts. Contact information is supplied to the SEC only through the request user-agent environment variable and is not embedded in analytical data.

\subsection{Filing inclusion rules}

The universe includes forms 10-K and 10-Q and excludes amendments. Accession number is the filing key. Fiscal-period end determines policy timing. Filing timestamp determines the beginning of observable event-day traffic. Missing report dates, missing primary documents, duplicate accessions, and filings outside the declared window are excluded by construction.

\subsection{Public-float selection algorithm}

For each CIK, the algorithm retrieves all USD-valued \texttt{EntityPublicFloat} facts attached to 10-K or 10-Q submissions. It keeps measurement dates in the predetermined interval and plausible values from \$0.5 to \$20 billion. For the latest measurement date, the modal repeated value is selected; ties use the earliest carrying filing. The analytical range is then restricted to \$2.5--10 billion and firms must have at least two filings on each side of the policy date.

\section{EDGAR Log Construction}

\subsection{Request validation}

A request must return a code below 300, have crawler and index flags equal to zero, and contain nonmissing IP and accession fields. The daily request count used for robot classification is computed over all valid EDGAR requests, not only sample accessions. This prevents a high-volume user from appearing low volume merely because few of its requests concern sample firms.

\subsection{Timing and deduplication}

Event day zero begins at the acceptance time stored in submissions metadata. Lexicographic time comparison is valid because both fields use zero-padded 24-hour time. Days one through seven are full calendar days. Sets of IPs are maintained separately for any asset, the primary document, and XML/XSD. The final filing-level union prevents one IP returning on eight days from being counted as eight distinct users.

\subsection{Zero versus missing}

A zero is written only when the corresponding daily log exists and contains no qualifying request. Every required date has an available log; hence no missing day is silently coded as zero. Daily log quality tables retain raw rows, valid rows, unique IP counts, robot counts, and matched target requests.

\section{Additional Results}

\subsection{Historical design and dynamic diagnostics}

\begin{figure}[H]
\centering
\includegraphics[width=.78\textwidth]{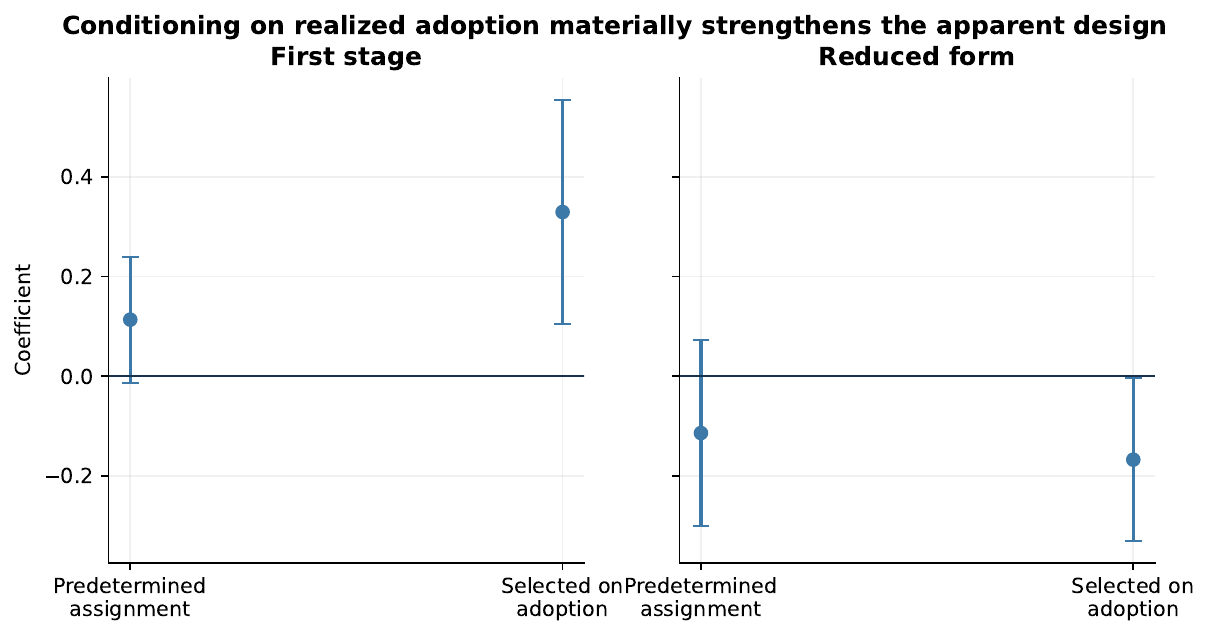}
\caption{Design sensitivity to conditioning on realized adoption. Bars are 95 percent firm-clustered confidence intervals.}
\label{fig:selection}
\end{figure}

\begin{figure}[H]
\centering
\begin{subfigure}{.49\textwidth}
\includegraphics[width=\textwidth]{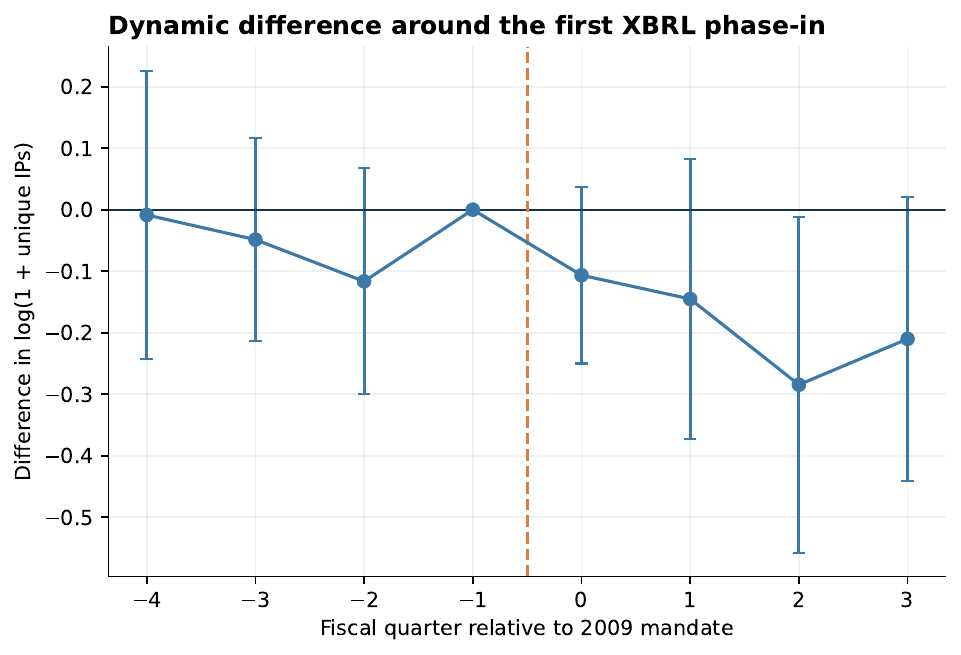}
\caption{Quarter-relative coefficients.}
\label{fig:eventstudy}
\end{subfigure}\hfill
\begin{subfigure}{.49\textwidth}
\includegraphics[width=\textwidth]{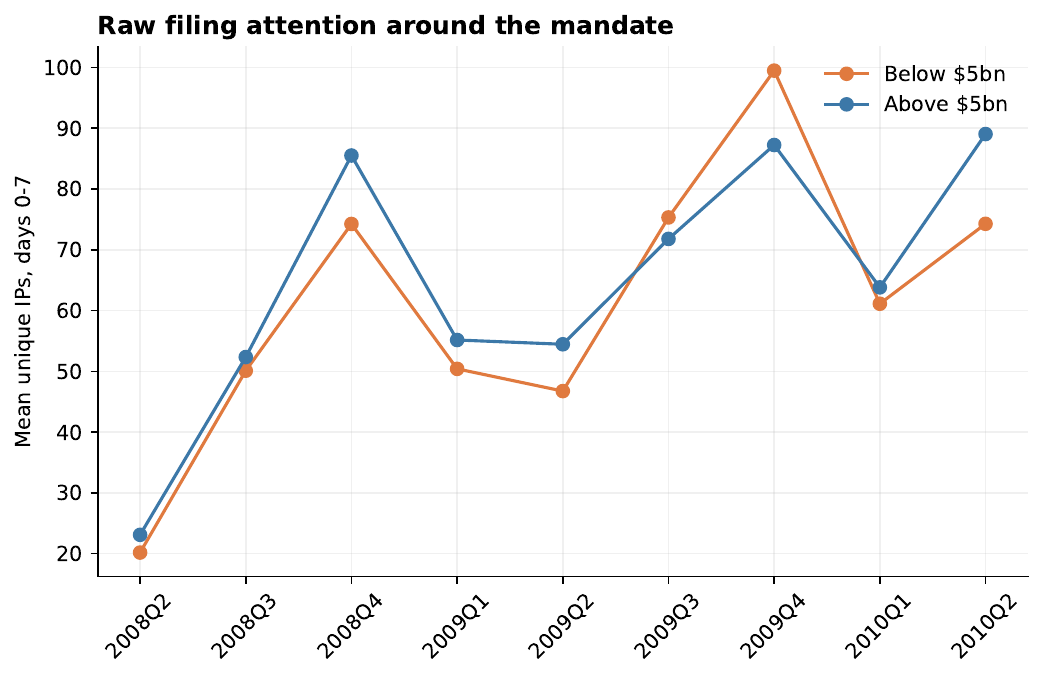}
\caption{Unadjusted quarterly means.}
\label{fig:rawtrends}
\end{subfigure}
\caption{Historical attention dynamics. The event-study bars are 95 percent firm-clustered confidence intervals; raw means are not regression adjusted.}
\end{figure}

\begin{figure}[H]
\centering
\begin{subfigure}{.49\textwidth}
\includegraphics[width=\textwidth]{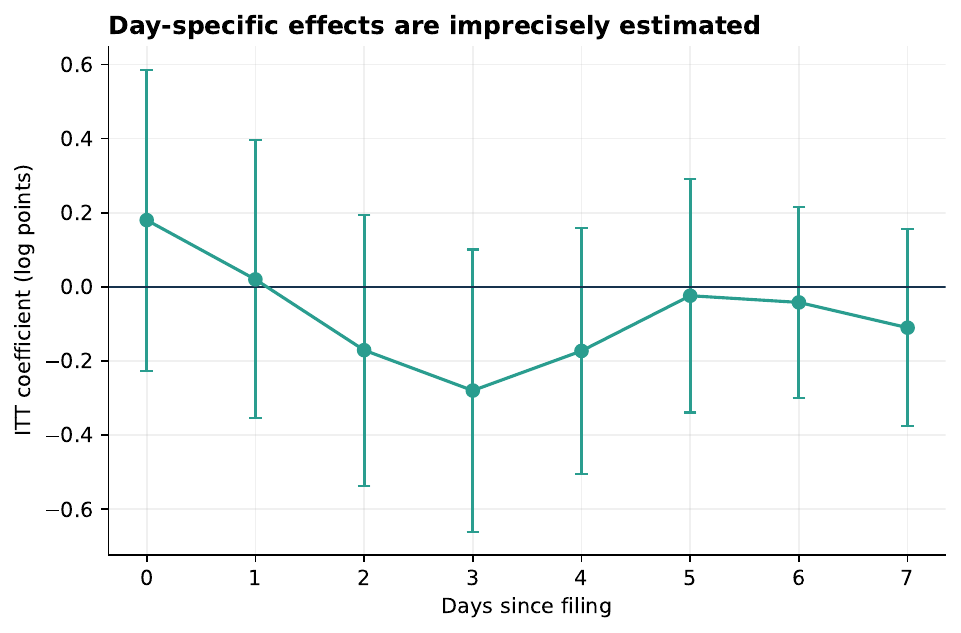}
\caption{Day-specific effects.}
\label{fig:days}
\end{subfigure}\hfill
\begin{subfigure}{.49\textwidth}
\includegraphics[width=\textwidth]{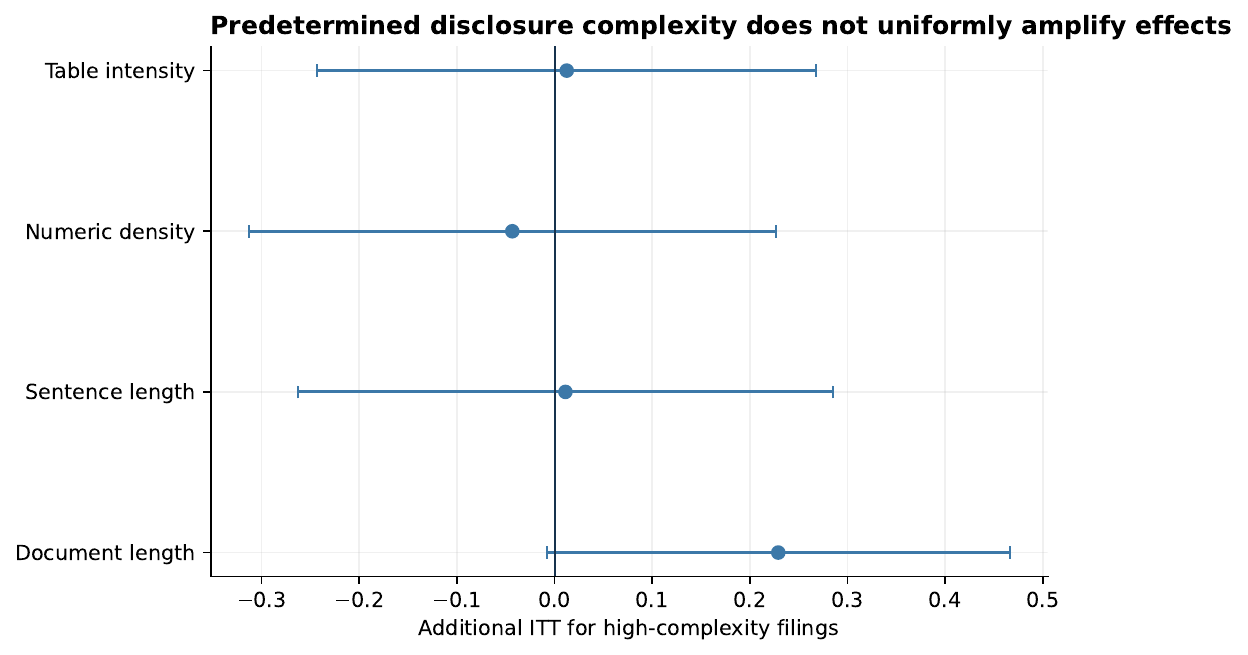}
\caption{Predetermined complexity interactions.}
\label{fig:heterogeneity}
\end{subfigure}
\caption{Historical timing and heterogeneity diagnostics. Points are intent-to-treat estimates and bars are 95 percent firm-clustered confidence intervals.}
\end{figure}

\subsection{Machine-readable result files}

The repository reports the complete coefficient vector for main outcomes, first stage, channel shares, robot alternatives, bandwidths, placebo dates, event quarters, event days, complexity interactions, and covariate continuity. Each record contains the estimate, standard error, raw $p$-value, confidence interval, filing count, and firm count. Outcome-family $q$-values are computed in code.

\subsection{Why no treatment-on-the-treated headline}

A Wald ratio would divide the reduced form by the estimated first stage. With a modest first stage, this ratio is unstable. More importantly, the exclusion restriction requires threshold assignment to affect traffic only through observed XBRL availability. Filer classification, compliance preparation, and correlated reporting technology could violate that condition. The reduced-form ITT is therefore the primary estimand.

\subsection{Interpreting XML requests}

The XML/XSD category includes instance documents and schemas. It is narrower than all EX-101 components and broader than a single economic task. A request shows retrieval, not parsing success. Nonetheless, the category is mechanically distinct from the primary HTML filename and transparently reproducible from the extension field.

\section{Modern-Module Construction and Validation}

\subsection{Bulk-archive census}

The modern census script scans every JSON member in the SEC submissions bulk archive, including recent histories stored in issuer files and older histories stored in supplementary members. It retains exact Forms 10-K and 10-Q between the declared dates and uses a CIK--accession primary key to remove overlap. Company name and SIC come from issuer-level metadata where available. The output is a 25MB CSV backed by a SQLite database, which permits independent count and uniqueness checks.

The Company Facts pass scans each namespace, concept, unit, and observation for CIKs in the filing census. An observation is assigned to a filing only by exact accession match. Counts include repeated concept observations for different contexts and units; distinct-concept coverage deduplicates namespace--concept pairs within accession. 
\texttt{EntityPublicFloat} is retained only when its value is numeric and attached to the same accession. The match rate to at least one standardized observation is \ModernFactCoverage{} percent.

\subsection{Sampling algorithm}

The diagnostic sample is generated from the census without consulting document text. Eligibility requires a 10-K, a nonmissing primary-document filename, a strictly positive same-accession public float, and a filing year in \{2018, 2020, 2022, 2024\}. When an issuer has multiple eligible annual reports in a filing year, the latest deterministic accession is retained. Within year, public float is ranked with unique tie breaks and split into four equal-sized groups. A NumPy random generator with seed 250707037 selects 25 rows per cell without replacement.

The chosen sample contains 400 reports and 394 distinct issuers. Six issuers appear in more than one sample year; no inference treats the rows as independent causal assignments. All regression uncertainty is HC1 rather than issuer-clustered because the repeated-issuer count is too small for reliable clustering; clustering by CIK produces substantively similar scale rankings but is not used as a precision claim.

\subsection{Text extraction and parser diagnostics}

Each primary-document URL is constructed from integer CIK, compact accession, and the SEC-reported filename. Downloads use a contact-bearing user agent, four polite workers, retry with exponential backoff, and a pause after successful retrieval. The failures file is empty. Total raw size is approximately 1.3GB.

Documents are decoded by the lxml HTML parser. Script and style nodes are removed. Visible whitespace is collapsed before regular-expression tokenization. Words contain alphabetic characters with optional apostrophes; numeric tokens allow signs, currency symbols, grouping punctuation, and percentages. Sentence counts use terminal punctuation and are explicitly labeled heuristic. Inline tag counts select names in the \texttt{ix:} namespace. A validation cross-tab shows zero reports flagged as Inline XBRL with zero detected Inline tags in the diagnostic sample.

\subsection{Index sensitivity and interpretation}

Within-year standardization prevents the adoption phase-in from mechanically determining cross-sectional index values. Log transforms limit domination by very large tag and word counts. Equal-weight means make the sign and contribution of every component auditable. The results are not driven only by composites: in Equation \ref{eq:modernscale}, Q4--Q1 gaps are positive for log fact observations, log concepts, and log Inline tags; human-burden primitives also rise with scale.

The interface-divergence index should not be compared cardinally across years because each component is standardized within year. Its purpose is to identify within-regime filings whose human and machine interfaces point in different directions. A task-specific alternative could weight facts relevant to valuation, narrative retrieval accuracy, or accessibility for assistive technology. Those are future validation targets.

\section{Paired Experiment and Counterfactual Protocol}

\subsection{Question construction and entity integrity}

Question generation reads SEC Company Facts by CIK and matches observations by exact accession. Instant concepts must end on the report date and have no start date. Duration concepts must end on that date and span 300--430 days; the duration nearest 365 days is preferred. Repeated contexts use the modal numeric value. Candidate concepts are Assets, Liabilities, AssetsCurrent, LiabilitiesCurrent, NetIncomeLoss or ProfitLoss, and three declared revenue alternatives. CIK and accession jointly identify the entity, preventing a corporation and operating partnership that share an accession from exchanging facts.

The 432-question pool is crossed with all twelve interfaces; no treatment assignment changes question composition. Text retrievers use visible filing text after scripts and styles are removed. BM25 and TF--IDF query terms are built from question language and declared target labels. Table-aware retrieval preserves table row and neighboring-cell structure. Oracle interfaces are generated from the accession-matched facts required for the answer. Real same-filing facts provide distractors; 421 of 432 padded pairs are within 30 words of their matched text budget, and the remaining 11 are disclosed rather than silently filled with synthetic prose.

\subsection{Evidence coverage and scoring audit}

Each task stores allowed evidence identifiers and one required support group per input fact. A citation passes only if cited blocks collectively intersect every required group. For HTML, a block joins a support group only when the target label or concept and its numerical value co-occur. This rule distinguishes a valid identifier from valid support. Before dispatch, required-fact coverage is 82.4 percent for BM25 HTML, 83.3 percent for TF--IDF reranking, 83.6 percent for table-aware HTML, 97.7 percent for retrieved XBRL, 99.8 percent for hybrid, and 100 percent for oracle, full-XBRL, and flat-JSON arms. A missing input fact makes grounded success impossible even if the model guesses the number.

Engineering connectivity responses and an initial permissive-citation run are archived outside the formal response directory and never enter a paper statistic. The formal audit verifies exact task-set equality for every model, removes duplicate retry records by model and cell identifier, rejects failed requests, and requires a Boolean grounded score. The full raw response remains available for independent regrading.

\subsection{API implementation and provenance}

All hosted implementations receive the same prompt, temperature zero, top-$k$ one, top-$p$ one, and deterministic response schema. Concurrent dispatch changes wall-clock order but not task content. Authentication is read only at runtime from a private configuration. Credentials are never copied into scripts, analytical outputs, provenance records, or the submission source bundle. For every request we retain requested and returned model identifiers, response ID, creation time, finish reason, usage, latency, retry count, dispatch order, raw text, parsed object, and grading fields. The submission-facing provenance records use provider-neutral model labels and include decoding parameters, timeout, worker count, random seed, and task-file digest; deployment routing is retained separately in the non-public audit record.

\begin{sloppypar}
The six completed implementations are Gemini 2.5 Flash, Gemini 2.5 Pro, DeepSeek V3.2, Qwen 3.7 Plus, Qwen3-235B-A22B, and Doubao Seed 1.8. These are provider-neutral model labels for specific hosted implementations, not claims about every deployment carrying the same family name. GLM-5 produced 4,215 successful main-experiment cells before persistent service nonresponse; its entire incomplete run is archived outside the formal response directory and contributes no observation. Kimi-K2 failed the connectivity gate. These exclusions are at the implementation level, never selected question by question. Deployment-level output variation remains part of external validity rather than an error term that pooling is allowed to conceal.
\end{sloppypar}

\subsection{Counterfactual transformation and exclusion rule}

The counterfactual builder draws a task-stratified 100-question subset using a fixed seed. Each needed value is deterministically transformed and calculated answers are recomputed. Original values are inserted as prior-period decoys; additional wrong-unit and segment-scale values test whether a model selects the correctly scoped evidence. The procedure creates evidence packets, not complete accounting-consistent synthetic filings. One generated liabilities-to-assets packet implied a ratio above 6,000 percent and repeatedly triggered provider nonresponse. The final rule excludes that question in all five conditions and all six models, leaving a fully paired 99-question analysis. No exclusion depends on whether a returned answer was correct.

\subsection{Inference and machine-readable outputs}

For each interface contrast, correctness is differenced within question and model. The pooled estimate first averages implementation-specific differences within question. Inference then clusters by the 97 unique accessions represented by the 432 questions, preserving pairing while allowing arbitrary dependence among questions from the same filing. Filing-level bootstrap intervals resample accessions with all their questions for 10,000 deterministic draws. Model-specific estimates use the same accession-clustered procedure and are reported without assuming six hosted systems form a random sample of all future models. Cross-model standard deviations and a full heat map show heterogeneity. CSV outputs contain model--interface summaries, accession-clustered paired contrasts, filing-bootstrap intervals, post hoc practical-equivalence tests, task interactions, size interactions, counterfactual contrasts, and filing-level construct-validation data. PDF and PNG renderings of every generated figure are produced from those tables rather than manually redrawn.

\section{Claim-to-Evidence Map}

\begin{longtable}{>{\raggedright\arraybackslash}p{.25\textwidth}>{\raggedright\arraybackslash}p{.31\textwidth}>{\raggedright\arraybackslash}p{.32\textwidth}}
\toprule
Claim & Supporting evidence & Explicit boundary \\
\midrule
\endfirsthead
\toprule
Claim & Supporting evidence & Explicit boundary \\
\midrule
\endhead
XBRL changes observed interface choice & Filing-level ITT estimates for primary HTML and XML/XSD requests & Modest first stage; individual channel estimates are imprecise \\
Direct traffic is not invariant to filtering & Same filings re-estimated under 50-request, 1,000-request, and SEC-flag screens & Filters do not identify true beneficial users \\
Modern filing architecture is joint human--machine & Census Inline-XBRL indicators for \ModernCensusFilings{} filings & Some forms and filers remain outside universal coverage \\
Human burden and machine accessibility are distinct & Primitive rank correlations and two-axis document plot & Indices are task-neutral diagnostics, not comprehension scores \\
Machine-ready coverage varies with issuer scale & Balanced-sample quartile gradients with year and SIC effects & Descriptive infrastructure disparity, not causal welfare inequality \\
AI can create an observability wedge & Model of intermediary service multipliers plus interface-sensitive traffic & Downstream AI use and service multipliers are not directly observed \\
Evidence addressability changes AI reliability & 432 standardized numerical questions crossed with twelve interfaces and six hosted model implementations, with strict SEC-grounded scoring and failure decomposition & One numerical task environment and response contract; no narrative synthesis or investor decisions \\
Machine-accessibility primitives have initial criterion validity & Issuer-grouped held-out prediction, filing-bootstrap intervals, interface-family tests, and leave-one-model-out checks & Validity is strongest for HTML-facing performance and is not a universal latent trait \\
\bottomrule
\end{longtable}

This map prevents conceptual mechanisms, historical reduced forms, modern descriptions, and randomized model outputs from being blended into a single causal claim. Each row can be strengthened only by adding evidence that crosses its stated boundary, not by relabeling the estimand.

\bibliographystyle{plainnat}
\bibliography{references,references_v2}

\end{document}